\documentclass[11pt,a4paper]{article}
\pdfoutput=1
\usepackage{./style/jheppub}

\usepackage{epsfig,multicol,bbm}

\usepackage{amsmath, amsthm, amssymb, bm, mathrsfs}
\usepackage[hyperpageref]{backref}

\title{Potential theory, path integrals and the Laplacian of the indicator}
\author{Rutger-Jan Lange}
\affiliation{University of Cambridge,\\
792 King's College, Cambridge, CB2 1ST, United Kingdom}
\emailAdd{rutger-jan.lange@cantab.net}
\abstract{
\noindent This paper links the field of potential theory --- i.e. the Dirichlet and Neumann problems for the heat and Laplace equation --- to that of the Feynman path integral, by postulating the following seemingly ill-defined potential:
\begin{equation*}
V(x):=\mp \frac{\sigma^2}{2}\,\nabla_{{x}}^2 \mathbbm{1}_{{x}\in D} 
\end{equation*}
\noindent where the volatility is the reciprocal of the mass (i.e. $m=1/\sigma^2$) and  $\hbar=1$. The Laplacian of the indicator can be interpreted using the theory of distributions: it is the $d$-dimensional analogue of the Dirac $\delta'$-function, which can formally be defined as $\partial_{x}^2 \mathbbm{1}_{x>0}$. 

\noindent We show, first, that the path integral's perturbation series (or Born series) matches the classical single and double boundary layer series of potential theory, thereby connecting two hitherto unrelated fields. Second, we show that the perturbation series is valid for all domains $D$ that allow Green's theorem (i.e. with a finite number of corners, edges and cusps), thereby expanding the classical applicability of boundary layers. Third, we show that the minus (plus) in the potential holds for the Dirichlet (Neumann) boundary condition; showing for the first time a particularly close connection between these two classical problems. Fourth, we demonstrate that the perturbation series of the path integral converges as follows:
\begin{center}
\begin{tabular}
{|l|c|r|}
\hline
\textit{mode of convergence} & absorbed propagator & reflected propagator \\\hline
convex domain & \textit{alternating} & \textit{monotone} \\ \hline
concave domain & \textit{monotone} & \textit{alternating} \\
\hline
\end{tabular}\end{center}
We also discuss the third boundary problem (which poses Robin boundary conditions) and discuss an extension to moving domains.}


\begin{document}
\maketitle


\setcounter{section}{0}
\setcounter{subsection}{0}
\setcounter{equation}{0}


\numberwithin{equation}{section}
\theoremstyle{plain}
\newtheorem{absorbed}{Proposition}
\newtheorem{reflected}[absorbed]{Proposition}
\newtheorem{absorbedlemma}{Lemma}
\newtheorem{reflectedlemma}[absorbedlemma]{Lemma}
\newtheorem{arrowlemma}[absorbedlemma]{Lemma}
\newtheorem{AbsorbedSeries}[absorbed]{Proposition}
\newtheorem{ReflectedSeries}[absorbed]{Propostion}
\newtheorem{CorollaryAbsorbed}{Corollary}
\newtheorem{Potential}[absorbed]{Propostion}
\newtheorem{SingularPotential}[absorbed]{Propostion}
\newtheorem{Theorem1}{Theorem}


\section{Introduction}

This paper links the field of potential theory --- i.e. the Dirichlet and Neumann problems for the heat and Laplace equations --- to that of the Feynman path integral, by postulating the following seemingly ill-defined potential:
\begin{equation*}
V(x):=\mp \frac{\sigma^2}{2}\,\nabla_{{x}}^2 \mathbbm{1}_{{x}\in D} 
\end{equation*}
where the Laplacian of the indicator can be interpreted using the theory of distributions. This is important for three reasons:
\begin{enumerate}
	\item Although potential theory was introduced by Green as early as 1828, \cite{Green1828}, this paper shows for the first time that single and double boundary layers are equivalent and follow directly from the first- and last-passage decompositions of the Brownian path.  
	\item Path integrals were introduced by Feynman in 1948 \cite{Feynman1948}, but the difficulty of incorporating boundary conditions has persisted. This paper shows how to impose absorbing, reflecting or elastic boundary conditions in $d\geq1$ dimensions, by using singular potentials such as above.
	\item The Dirac $\delta$-function and its derivative $\delta'(x)$ have been known at least since Dirac's seminal work \cite{Dirac1930} of 1930. Both the one-dimensional version and the multi-dimensional generalisations, as they are usually made, are only non-zero at a single point. However, a different generalisation is possible. A point in one dimension can be considered as the boundary of a halfline, and the Dirac $\delta$-function and its derivative can formally be viewed as $\partial_x  \mathbbm{1}_{{x}>0}$ (the inward derivative of the indicator) and $\partial_x^2  \mathbbm{1}_{{x}>0}$ (the Laplacian of the indicator). The latter view is taken in this paper. This leads to the multidimensional versions $-n_x\cdot\nabla_x\mathbbm{1}_{{x}\in D}$ and $\nabla_{{x}}^2 \mathbbm{1}_{{x}\in D}$, respectively, which are supported by surfaces rather than points. Both quantities have --- to the author's best knowledge --- not formally been defined before. Apart from their use in this paper, we suspect that they may have further, independent utility. 
\end{enumerate}
For practical purposes we shall approximate the indicator $\mathbbm{1}_{{x}\in D}$ by a \textit{bump function} that we shall indicate by $I_\epsilon(x)$, which is smooth\footnote{At least in the direction normal to the boundary, if the boundary is piecewise smooth.} for all $\epsilon>0$, and approximates the indicator \textit{from below}, i.e.
\begin{equation}
I_\epsilon(x)\geq 0 \quad \quad I_\epsilon(x)\leq \mathbbm{1}_{{x}\in D}\quad\quad \underset{\epsilon \searrow 0}\lim\; I_\epsilon(x)=\mathbbm{1}_{x\in D}
\end{equation} 
It will become clear that this choice is the natural one to make. This introduction discusses these three contributions sequentially.

\subsection{Potential theory}

Potential theory calls for the construction of harmonic functions, i.e. satisfying $\nabla^2 f = 0$ in $D$, with the further condition that they satisfy certain boundary conditions at $\partial D$. The Dirichlet problem prescribes the value at $\partial D$, while the Neumann problem prescribes the normal derivative at $\partial D$. Green \cite{Green1828} realised as early as 1828 that the problem can be reduced to finding the Green function, as it is now known. Furthermore, he realised that nature solves the problem: an electric charge at $x$, placed inside a perfect conductor, causes an electric potential at $y$ that is equal to the Green function of the Dirichlet problem. 

It turns out that the classical Dirichlet problem is not solvable for geometries with isolated boundary points (see \cite{Zaremba1911}) or sharp thorns (see \cite{Lebesgue1913}). But the \textit{modified} Dirichlet problem is well-defined: it only asks for the boundary conditions to be satisfied at all \textit{regular} boundary points (see e.g. \cite{PortStone1978}). Physicists never worried about such peculiar cases, because nature solves the modified problem: an induced charge density on the conductor \textit{will} exist, even if the conductor is shaped like a thorn. At irregular boundary points multiple `normal' directions exist, and an electrical force acting in any of those is allowed. 

Kakutani \cite{Kakutani1944} realised that Brownian motion can be used to solve the modified Dirichlet problem. The solution at $x$ can be obtained by 1) simulating many Brownian motions starting at $x$ until they hit $\partial D$, by 2) assigning to each path the supposed boundary value at its first passage over $\partial D$, and by 3) taking an expectation over all paths. This prescription solves the modified Dirichlet problem, because, as Chung \cite[p.~54]{Chung1995} notes:
\begin{quote}
although there may be irregular points on $\partial{D}$, almost no path will ever hit them. Thus they are not really there so far as the paths are concerned. 
\end{quote}

Brosambler's \cite{Brosamler1976} discovery, that reflected (rather than absorbed) Brownian motion could reproduce the solution to the Neumann problem, further strengthened the case for the use of stochastic processes to study partial differential equations. Elastic Brownian motion (which is either absorbed or reflected each time it hits the boundary) turned out to be useful for the third boundary value problem, which poses Robin boundary conditions. 

Subsequent literature (e.g. \cite{BalianBloch1974,BalianDuplantier1977,BalianDuplantier1978,HanssonJaffe1983a,HanssonJaffe1983b,Bordag1999,Bordag2001,Bordag2002,Bordag2005,Maghrebi2011}) followed the now classical approach by Balian and Bloch \cite{BalianBloch1970}: to find the Green function for the problem, the \textit{ansatz} of a double (single) boundary layer is made for the Dirichlet (Neumann) problem. In contrast, we show that 1) neither single nor double boundary layers need to be based on an ansatz, but, in fact, follow from the first- and last-passage decompositions of Brownian paths, 2) either problem may be solved with either method and their distinction thus is arbitrary, and 3) boundary layers may be used for irregular as well as regular domains. The literature above has only considered smooth domains, again by following the example of \cite{BalianBloch1970}. 

The extension from smooth domains to piecewise smooth domains may seem of only minor relevance. But the following question, since being raised by Kac \cite{Kac1966}, has received much attention (e.g. \cite{Stewartson1971,Protter1987,Giraud2010}): if all the eigenvalues of the Dirichlet (or Neumann) solution are given, can one uniquely reconstruct the domain? It turns out that the answer is `yes' if the domain is smooth and `no' if sharp corners are allowed. We merely require that the domain $D$ allows Green's second identity (allowing edges, corners and cusps), and thus we provide a tool for calculating the Green function for domains of either type.  

\subsection{Path integrals}

Feynman \cite{Feynman1948} developed path integrals to describe the movement of a quantum particle under the influence of a potential $V$. Kac gave the probabilistic interpretation in \cite{Kac1949} and \cite{Kac1951}: a Brownian particle moves freely, but may be annihilated by a positive potential $V$. The probability that this happens, at any location, equals the product of the strength of the potential at that location and the (infinitesimal) amount of time spent there. And a negative potential creates particles, again at a rate corresponding to its magnitude.

But there are several problems regarding path integrals. First, path integrals can only be calculated exactly very occasionally, although perturbation series (or Born series) can be easily written down; see e.g. \cite[p.~128]{FeynmanHibbs1965} or \cite[p.~161]{Ryder1996}.
Second, the treatment of even the simplest boundary-value problems is notoriously complicated within the path integral framework. Kleinert, for example, writes in \cite{Kleinert1979}:

\begin{quote}
Considering the present widespread use of path integrals [...], it is surprising how many standard text book problems of quantum mechanics have not been solved within this framework. [...] In this note we would like to exhibit the path integration for the particle in a box (infinite square well). While in Schr\"{o}dinger theory this system has a trivial solution, a careful classification of paths is needed before Feynman's formula can be evaluated.
\end{quote}
The paper then shows how to evaluate Feynman's formula for the one-dimensional particle in a box. No progress, however, has been made in evaluating Feynman's formula for bounded domains in $d\geq 1$. Boundary-value problems confine the particle to a particular region of space, but the Gaussian integrals are much easier, at least analytically, if they stretch across the whole real line. 

Even in the one-dimensional case discussed above, it is tempting to postulate an infinite potential outside of the box. The interpretation is that of an infinite annihilation rate, such that every Brownian path spending even a small time outside the box is annihilated. We could take $V(x)=\mathbbm{1}_{x \notin D}$ and let the coupling constant $\lambda \to \infty$, so that the annihilation rate outside the box goes to infinity. But, if we let $\lambda \to \infty$, then all terms in the perturbation series (or Born series) become infinite. Even though the series formally still converges, this obviously diminishes its practicality. To overcome this problem, the use of Dirac $\delta$-function potentials has been suggested, which are infinite at the edge of the box but zero beyond. Various authors (e.g. \cite{Clark1980,Albeverio1984,Lawande1988,Albeverio1994,Grosche1990}) have considered these so-called \textit{point interactions}. All perturbation terms are now finite and the series thus converges in a meaningful manner, but the Dirac $\delta$-potential is not strong enough to confine the particle. It is a textbook result that a particle can tunnel through a Dirac $\delta$-potential. Again we could let $\lambda \to \infty$, but then all perturbation terms become infinite. While we already \textit{know} the solution for the one-dimensional particle in a box, it is not clear what potential manages to 1) replicate the solution and 2) allow for a meaningful perturbation series. 

A further problem with point interactions is that how to generalise them to treat higher-dimensional boundary problems is not obvious. There seems to be scarce literature on what we may call \textit{surface interactions}. One purported reason is that treating singular potentials is hard, even in $d=1$.

\begin{figure}[t]
\includegraphics[width=\textwidth]{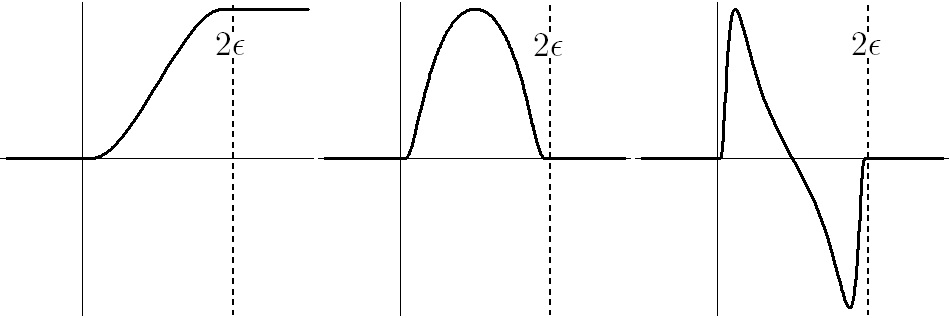}
\caption{This graph shows $I_\epsilon(x)$, $I'_\epsilon(x)$ and $I''_\epsilon(x)$ when the domain is taken to be the positive real line. As in \cite[p.~206]{KaratzasShreve1991}, we have chosen $I'_\epsilon(x)$ as the $C^\infty$ function $I'_\epsilon(x)=c/\epsilon \exp\left(1/([1-x/\epsilon]^2-1)\right)$ for $x$ between $0$ and $2 \epsilon$, and zero otherwise. The constant $c$ is chosen such that its integral equals $1$. For our intuition we can think of a positive (negative) potential as destroying (creating) particles. This helps to explain why the potential $\sigma^2/2\,I_\epsilon''(x)$, as in the rightmost graph, is absorbing from the left and reflecting from the right.}
\label{heaviside}
\end{figure}

\subsection{Intuition for the Laplacian of the indicator}

\begin{figure}[t]
\includegraphics[width=\textwidth]{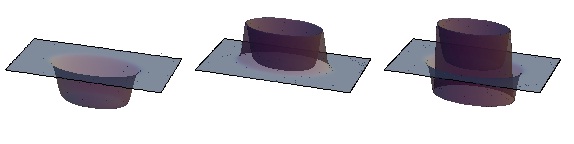}
\caption{While for any $\epsilon>0$ the negative bump function is continuously differentiable to all orders, for $\epsilon \searrow 0$ we get $-\mathbbm{1}_{x \in D}$, $-n \cdot \nabla_x \mathbbm{1}_{x \in D}$ and $-\nabla_x^2 \mathbbm{1}_{x \in D}$. For our intuition we can think of the rightmost graph as resembling an elliptical castle, with a moat in front of the castle walls. The wall annihilates particles while the moat creates them, such that the castle reflects Brownian particles that approach it from the outside. Particles that approach the wall from the inside are annihilated.}
\label{castle}
\end{figure}

In this paper we show that a Brownian motion that is absorbed (reflected) at $\partial D$ is consistent with a path integral formulation when the particle is allowed in all of $\mathbb{R}^d$ but is acted upon by a potential $V$, where the potential is proportional to minus (plus) the Laplacian of the indicator of some domain $D$. For practical purposes, we approximate the indicator using a bump function that is identically zero outside of $D$: see e.g. the one-dimensional example in Figure \ref{heaviside}, where the domain is taken to be the positive real line; and the two-dimensional example in Figure \ref{castle}, where the domain is taken to be an ellipse.

In terms of \textit{why} this potential does the job, we can say the following. The second derivative of the indicator has `higher' peaks than the first derivative. While the Dirac $\delta$-function is absolutely integrable, the Dirac $\delta'$-function is not. Therefore, the potential is indeed strong enough to contain the particle. 

But we have also noted that positive potentials destroy paths, while negative potentials create paths. Through the limiting procedure in Figures \ref{heaviside} and \ref{castle}, we see that the Laplacian of the indicator has a positive and a negative peak. The positive peak destroys a particle, while the negative peak creates one. If a Brownian particle approaches the potential and is first presented with the positive peak, then the Brownian particle is destroyed. If it is first presented with the negative peak, however, an extra particle is created. This extra particle is subsequently destroyed by the positive peak, and only the original particle remains: it is reflected off the boundary. While not overly rigorous, this intuition may help to explain why the potential for the absorbed and reflected propagators differs only by a sign.  

As a result, the $\sigma^2/2\,\delta'$-potential is reflecting from above. Thus, for $x>0$, we obtain the following propagation density:
\begin{equation}
\label{psireflected}
\psi(y,t|x,s)=\left\{
\begin{array}{l@{\hspace{2mm}}l}
\displaystyle B(y,t|x,s)+B(y,t|x^*,s)& x>0, y>0,\\[1ex]
\displaystyle 0& x>0, y<0,\\[1ex]
\end{array}
\right.
\end{equation}
The free-Brownian propagator from space-time coordinate $(x,s)$ to $(y,t)$ is denoted by  $B(y,t|x,s)$ (defined in subsection \ref{subsection2.1}), and $x^*$ is the mirror-coordinate of $x$, i.e. $x^*=-x$. When that starting point $x$ is above zero, the boundary at zero is reflecting and  the particle can never reach $y<0$. It is thus clear that $\psi$ is discontinuous in $y$, for $x>0$. The derivative, however, is continuous (zero on both sides). For the heat equation, which is of second order in the space variable, it is commonly understood that discontinuities in the derivative correspond to Dirac $\delta$-potentials, while discontinuities in the value correspond to Dirac $\delta'$-potentials. Our $\psi$ of \eqref{psireflected} could thus be expected to satisfy a partial differential equation involving a Dirac $\delta'$-function. But as it happens, the Dirac $\delta'$-potential has caused controversy. This began with an exchange of arguments between two parties \cite{Albeverio1988,Zhao1992,Albeverio1993}, and was resolved by an independent note  \cite{Griffiths1993}, or so it seemed. The more recent paper \cite{Coutinho1997} claims that even this resolution had its flaws.

In essence, the source of the controversy is the following: on the one hand, we know that the second derivative of a discontinuity --- in the distributional sense --- equals a Dirac $\delta'$-function. Often, the Dirac $\delta$-function is considered as the limit of a Gaussian centred at zero, or otherwise as the limit of a family of even distributions. In either case, the Dirac $\delta$-function is symmetric and, as emphasised in e.g. \cite{Griffiths1993}, it deals with discontinuous integrands $f$ as follows: 
\begin{equation}
\int_{-\epsilon}^\epsilon\,\delta(y)\,f(y)\;dy = \frac{1}{2} (f(0+)+f(0-)\quad\forall \epsilon > 0
\end{equation}
Although a Gaussian centred at zero is even, its derivative is odd. When $f$ is discontinuous across zero, therefore, the following quantity does \textit{not} exist: 
\begin{equation}
\int_{-\epsilon}^\epsilon\,\delta'(y)\,f(y)\;dy = \pm \infty \quad \forall \epsilon > 0
\end{equation}
This was also was noted by \cite[p.~3943]{Coutinho1997} and can be checked by direct calculation, or by realising that the `slope' of $f$ is $\pm \infty$ at zero when $f$ is discontinuous there. Therefore, we cannot claim that 1) $\psi$ is discontinuous across zero, as well as that 2) $\psi$ satisfies the Schr\"{o}dinger equation with a Dirac $\delta'$-potential. The latter involves the term $\delta'\,\psi$, which is undefined for discontinuous $\psi$. Still, it is clear that our $\psi$ of \eqref{psireflected} is discontinuous, so what PDE should it satisfy? In fact, a slightly different route to the problem turns out to be fruitful. Consider a smooth potential $V_\epsilon$ that is singular in the limit $\epsilon \searrow 0$. In section \ref{subsection3.3} we show that in such a case the following decompositions hold:
\begin{equation}
\label{decompositions}
\begin{array}{r@{\hspace{1mm}}c@{\hspace{1mm}}l}
\displaystyle\psi(y,t|x,s)&=&\displaystyle B(y,t|x,s)-\underset{\epsilon \searrow 0}\lim\;\underset{\hspace{-3mm}s}{\overset{\hspace{3mm}t}\int} d\tau \underset{\hspace{-3mm}-\infty}{\overset{\hspace{3mm}\infty}\int} d\alpha\;B(y,t|\alpha,\tau)\,V_{\epsilon}(\alpha)\,\psi(\alpha ,\tau |x,s),\\[2ex]
\displaystyle\psi(y,t|x,s)&=&\displaystyle B(y,t|x,s)-\underset{\epsilon \searrow 0}\lim\;\underset{\hspace{-3mm}s}{\overset{\hspace{3mm}t}\int} d\tau \underset{\hspace{-3mm}-\infty}{\overset{\hspace{3mm}\infty}\int} d\alpha\;\psi(y,t|\alpha,\tau)\,V_{\epsilon}(\alpha)\,B(\alpha ,\tau |x,s).
\end{array}
\end{equation} 
These decompositions are well-suited to obtaining series solutions: we can substitute the equations into themselves, substituting the definition of $\psi$, as given by the left-hand side, into $\psi$ on the right-hand side. This series solution is well-behaved for all smooth potentials, and may (or may not) be well-behaved for singular potentials. For the Dirac $\delta'$-potential, when viewed as the limit of the derivative of a Gaussian centred at zero, the series is not well-behaved: its second correction term is infinite. If, instead, we take the Dirac $\delta'$-function as in the rightmost graph in Figure \ref{heaviside}, then each term in the series is well-behaved. Therefore we take the potential $V_\epsilon(x)$ as follows:
\begin{equation}
\label{potential}
V_\epsilon(x)=\frac{\sigma^2}{2}I''_\epsilon(x),
\end{equation}
where the bump function $I_\epsilon(x)$ approaches the indicator $\mathbbm{1}_{x>0}$ from below, i.e. 
\begin{equation}
\label{bump function}
\displaystyle I_\epsilon(x)\geq 0\quad\quad I_\epsilon(x)\leq \mathbbm{1}_{x>0}\quad \quad\underset{\epsilon \searrow 0}\lim I_\epsilon(x)=\mathbbm{1}_{x>0},
\quad \mbox{$\forall \epsilon>0$}.
\end{equation}
One may check by direct calculation that $\psi$ of \eqref{psireflected} satisfies the decompositions \eqref{decompositions} when the potential is given by \eqref{potential}, and when both $x,y>0$. In fact, it does not matter how $\psi$ is defined for $y<0$: as long as $\psi$ is equal to the reflected density for $x,y>0$, the decompositions \eqref{decompositions} hold for $x,y>0$. This is because the potential \eqref{potential} only `feels' whatever is to the right of the origin. In fact, the absorbed density
\begin{equation}
\label{psiabsorbed}
\psi(y,t|x,s)=\left\{
\begin{array}{l@{\hspace{2mm}}l}
\displaystyle B(y,t|x,s)-B(y,t|x^*,s)& x>0, y>0,\\[1ex]
\displaystyle 0& x>0, y<0,\\[1ex]
\end{array}
\right.
\end{equation}
satisfies the decompositions \eqref{decompositions} with the potential
\begin{equation}
\label{potentialabsorbed}
V_\epsilon(\alpha)=-\frac{\sigma^2}{2}I''_\epsilon(x).
\end{equation}
for $x,y>0$. The solution $\psi$ of \eqref{psiabsorbed} is not even discontinuous across zero. Thus we conclude that the `right-handed' Dirac $\delta'$-potential, as defined by $I_\epsilon''$, does not necessarily lead to discontinuities in $\psi$, and can be used to incorporate both absorbing and reflecting boundary conditions. 

In both the reflecting and absorbing case, we may use the decompositions \eqref{decompositions} to obtain a series solution that is well-behaved. Substitution the first decomposition into itself, we obtain
\begin{equation*}
\label{decompositionexpansion}
\begin{array}{r@{\hspace{1mm}}c@{\hspace{1mm}}l}
\displaystyle\psi(y,t|x,s)&=&\displaystyle B(y,t|x,s)-\underset{\epsilon \searrow 0}\lim\;\underset{\hspace{-3mm}s}{\overset{\hspace{3mm}t}\int} d\tau \underset{\hspace{-3mm}-\infty}{\overset{\hspace{3mm}\infty}\int} d\alpha\;B(y,t|\alpha,\tau)\,V_{\epsilon}(\alpha)\,B(\alpha ,\tau |x,s)\\
&&\hspace{-2.3cm}+\displaystyle \underset{\epsilon \searrow 0}\lim\;\underset{\hspace{-3mm}s}{\overset{\hspace{3mm}t}\int} d\tau_2 \underset{\hspace{-3mm}-\infty}{\overset{\hspace{3mm}\infty}\int} d\alpha_2\;B(y,t|\alpha_2,\tau_2)\,V_{\epsilon}(\alpha_2)\,\left(\underset{\epsilon \searrow 0}\lim\;\underset{\hspace{-3mm}s}{\overset{\hspace{3mm}\tau_2}\int} d\tau_1 \underset{\hspace{-3mm}-\infty}{\overset{\hspace{3mm}\infty}\int} d\alpha_1\;B(\alpha_2,\tau_2|\alpha_1,\tau_1)\,V_{\epsilon}(\alpha_1)\,\psi(\alpha_1,\tau_1|x,s)\right)
\end{array}
\end{equation*}
The free term on the right-hand side remains for both the absorbed and reflected solutions $\psi$. Depending on the sign of the potential, the first correction term gives either $+B(y,t|x^*,s)$ or $-B(y,t|x^*,s)$, and the second correction term vanishes. We have thus obtained a consistent prescription: if we let $\epsilon \searrow 0$ in the potential $\pm\frac{\sigma^2}{2}I''_\epsilon(x)$, then the solution will approach $\psi$ of \eqref{psireflected} or \eqref{psiabsorbed} for all $x,y>0$. When the limit is reached, both solutions satisfy the decompositions \eqref{decompositions}. Lastly, the term $I''_\epsilon(y) \psi(y,t|x,s)$ exists, and can be integrated to give the result $-\partial_y\psi(y,t|x,s)|_{y=0+}$. 

In addition to treating boundaries that are purely absorbing or reflecting, we can also treat `elastic' boundaries. Elastic boundaries reflect the Brownian particle with probability $\kappa\,dt$ and absorb it with probability $(1-\kappa\,dt)$, when $dt$ is the infinitesimal time spent at the boundary and where $\kappa>0$. For an elastic boundary at zero for $d=1$ see e.g. \cite{Rajabpour2009} to find the following propagation density:
\begin{equation}
\label{psi2}
\psi(y,t|,x,s)=B(y,t|x,s)+B(y,t|x^*,s)-2 \kappa \int_0^\infty d\alpha\; e^{-\kappa \alpha}\,B(y+\alpha,t|x^*,s)\quad x>0,y>0
\end{equation}
This can be obtained by taking a reflecting Brownian motion and putting annihilating Dirac $\delta$-potential just above zero. We can confirm by direct calculation that $\psi$ of \eqref{psi2} satisfies the decompositions \eqref{decompositions} for $x,y>0$, when the potential is equal to 
\begin{equation}
\label{potential2}
\begin{array}{r@{\hspace{1mm}}c@{\hspace{1mm}}l}
V_\epsilon(x)&=&\displaystyle\frac{\sigma^2}{2}I''_\epsilon(x)+\kappa\,\sigma^2\, I'_\epsilon(x), \\
\end{array}
\end{equation}
and where the bump function is again given by \eqref{bump function}. It is clear, both from the solution \eqref{psi2} as well as from the potential \eqref{potential2}, that elastic Brownian motion becomes reflected Brownian motion as $\kappa \searrow 0$. What is \textit{not} clear from \eqref{potential2}, however, is that as $\kappa \nearrow \infty$ we obtain absorbed Brownian motion. We know by physical intuition that the boundary must become absorbing as $\kappa \nearrow \infty$. For the one dimensional case, we can do the integration in \eqref{psi2} for $\kappa \nearrow \infty$ and show that, in this limit, the elastic density goes to the absorbed density.  

As we have shown, the Dirac $\delta'$-interaction is non-trivial even in one dimension. We differ further from the literature on point interactions by treating the higher-dimensional analogues of the Dirac $\delta$- and $\delta'$-function as $-n\cdot\nabla_x\mathbbm{1}_{{x}\in D}$ and $\nabla_x^2 \mathbbm{1}_{x \in D}$, respectively. The elastic potential in $d$ dimensions, for example, can be written as  
\begin{equation}
\label{potential3}
\begin{array}{r@{\hspace{1mm}}c@{\hspace{1mm}}l}
V_\epsilon(x)&=&\displaystyle\frac{\sigma^2}{2}\nabla_x^2 I_\epsilon(x)-\kappa\,\sigma^2\, n_x\cdot\nabla_x I_\epsilon(x).\\
\end{array}
\end{equation}
Here $n_x$ can be defined to exist for all $x$, for example as the outward normal of the boundary point nearest to $x$. The potential scales with $\sigma^2$, because when paths are more volatile, potentials with small support need to grow in magnitude to achieve the same effect. 

The Laplacian of the indicator is supported by a surface, whereas the usual higher-dimensional generalisations of Dirac $\delta$-functions and its derivatives are supported by points. This generalisation is useful because surface interactions can lead to boundary conditions in $d \geq 1$, while point interactions cannot. Naturally, point and surface interactions coincide for $d=1$. 

This paper is organised as follows: section \ref{section2} discusses Brownian motion in the context of potential theory. It shows that the classical single and double boundary layers follow from first- and last-passage distributions of Brownian motion (i.e. they are equivalent), and that they are useful for irregular domains. Section \ref{section3} discusses Brownian motion in the context of path integrals. It derives first- and last-interaction decompositions in the presence of a (possibly singular) potential $V$. Section \ref{section4} shows that sections \ref{section2} and \ref{section3} can be unified if one postulates derivatives of the bump function as the potential. Finally section \ref{section5} shows how to extend this work to moving boundaries.

\section{Brownian motion and potential theory}
\label{section2}

\subsection{Brownian motion with boundary conditions}
\label{subsection2.1}

In $d$ dimensions, the transition density of a standard Brownian motion is as follows:
\begin{equation}
\label{BrownianDensity}
B(y,t|x,s)=\displaystyle\frac{1}{[2\pi(t-s)]^{d/2}}\;\mbox{\textit{\large e}}^{-\frac{\left|y-x\right|^2}{2 \sigma^2(t-s)}}
\end{equation}
where $B(y,t|x,s)$ is equal to the (marginal) probability that a Brownian particle moves from the `backward' space-time coordinate $(x,s)$ to the `forward' space-time coordinate $(y,t)$. Formally, Brownian motion is defined as a continuous process, with independent increments, such that the increment during $dt$ is normally distributed with mean zero and variance $dt$. The explicit representation \eqref{BrownianDensity} shows that the Brownian density $B$ satisfies
\begin{equation}
\label{BrownianMotion}
\begin{array}{rr@{\hspace{1mm}}c@{\hspace{1mm}}l}
\text{\scriptsize{forward PDE}}&\displaystyle\Big(\frac{\partial}{\partial t} - \frac{1}{2}\nabla_y^2\Big)B(y,t|x,s)&=&0\\[1.5ex]
\text{\scriptsize{backward PDE}}&\displaystyle\Big(\frac{\partial}{\partial s} + \frac{1}{2}\nabla_x^2\Big)B(y,t|x,s)&=&0\\[1.5ex]
\text{\scriptsize{forward STC}}&\displaystyle\lim_{s \nearrow t} B(y,t|x,s)&=&\delta(|y-x|)\\[1.5ex]
\text{\scriptsize{backward STC}}&\displaystyle\lim_{t \searrow s} B(y,t|x,s)&=&\delta(|y-x|)\\
\end{array}
\end{equation}
where PDE stands for partial differential equation and STC stands for short-time condition. The STCs are satisfied because in a short time the particle stays where it is, and the PDEs are satisfied because the transition density is unbiased, i.e.
\begin{eqnarray*}
B(y,t|x,s)=\mathbb{E}\,B(y-dB,t-dt|x,s)\\
B(y,t|x,s)=\mathbb{E}\,B(y,t|x+dB,s+ds)
\end{eqnarray*}
and using It\^{o}'s lemma gives both PDEs. 

The transition density of absorbed Brownian motion (ABM) in $D$ is indicated by $A(y,t|x,s)$ and satisfies the following set of equations:
\begin{equation}
\label{AbsorbedBrownianMotion}
\begin{array}{rr@{\hspace{1mm}}c@{\hspace{1mm}}lr}
\text{\scriptsize{forward PDE}}&\displaystyle\Big(\frac{\sigma^2}{2}\nabla_y^2-\frac{\partial}{\partial t} \Big)A(y,t|x,s)&=&0&\\[1.5ex]
\text{\scriptsize{backward PDE}}&\displaystyle\Big( \frac{\sigma^2}{2}\nabla_x^2+\frac{\partial}{\partial s}\Big)A(y,t|x,s)&=&0&\\[1.5ex]
\text{\scriptsize{forward BC}}&A(\beta,t|x,s)&=&0&\beta \in \partial D^r\\[1.5ex]
\text{\scriptsize{backward BC}}&A(y,t|\beta,s)&=&0&\beta \in \partial D^r\\[1.5ex]
\text{\scriptsize{forward STC}}&\displaystyle\underset{s \nearrow t}\lim A(y,t|x,s)&=&\delta(|y-x|)&\\[1.5ex]
\text{\scriptsize{backward STC}}&\displaystyle\underset{t \searrow s}\lim A(y,t|x,s)&=&\delta(|y-x|)&\\
\end{array}
\end{equation}
for $x,y \in D$. BC stands for boundary condition and the BC holds at all \textit{regular} boundary locations $\beta$, indicated by $\partial D^r$. The definition of a regular boundary point is one that allows a tangent plane. The BCs are satisfied because no Brownian particle can move to or from a regular boundary point without being absorbed. We do not need to specify what happens at irregular boundary points, because `almost no path will ever hit them'   \cite[p.~54]{Chung1995}).

The transition density of reflected Brownian motion (RBM) is indicated by $R(y,t|x,s)$ and satisfies the following set of equations:
\begin{equation}
\label{ReflectedBrownianMotion}
\begin{array}{rr@{\hspace{1mm}}c@{\hspace{1mm}}lr}
\text{\scriptsize{forward PDE}}&\displaystyle\Big(\frac{\sigma^2}{2}\nabla_y^2-\frac{\partial}{\partial t} \Big)R(y,t|x,s)&=&0&\\[1.5ex]
\text{\scriptsize{backward PDE}}&\displaystyle\Big( \frac{\sigma^2}{2}\nabla_x^2+\frac{\partial}{\partial s}\Big)R(y,t|x,s)&=&0&\\[1.5ex]
\text{\scriptsize{forward BC}}&\displaystyle n_\beta \cdot \overrightarrow{\nabla}_\beta R(\beta,t|x,s)&=&0& \beta \in \partial D^r\\[1.5ex]
\text{\scriptsize{backward BC}}&\displaystyle R(y,t|\beta,s) \overleftarrow{\nabla}_\beta \cdot n_\beta &=&0& \beta \in \partial D^r\\[1.5ex]
\text{\scriptsize{forward STC}}&\displaystyle\lim_{s \nearrow t} R(y,t|x,s)&=&\delta(|y-x|)&\\[1.5ex]
\text{\scriptsize{backward STC}}&\displaystyle\lim_{t \searrow s} R(y,t|x,s)&=&\delta(|y-x|)&\\
\end{array}
\end{equation}
where the BCs are satisfied because a Brownian particle is reflected in the normal direction, at any regular boundary point $\beta$. As a result $R(y,t|\beta,s)$ and $R(y,t|\beta+\epsilon,s)$ are equal to first order in $\epsilon$, if $\beta$ is a regular boundary coordinate and $\epsilon$ is small displacement in the normal direction. Again we do not specify what happens at irregular boundary points.

The transition density of elastic Brownian motion (EBM) is indicated by $E(y,t|x,s)$ and satisfies the following set of equations:
\begin{equation}
\label{ElasticBrownianMotion}
\begin{array}{rr@{\hspace{1mm}}c@{\hspace{1mm}}lr}
\text{\scriptsize{forward PDE}}&\displaystyle\Big(\frac{\sigma^2}{2}\nabla_y^2-\frac{\partial}{\partial t} \Big)E(y,t|x,s)&=&0&\\[1.5ex]
\text{\scriptsize{backward PDE}}&\displaystyle\Big( \frac{\sigma^2}{2}\nabla_x^2+\frac{\partial}{\partial s}\Big)E(y,t|x,s)&=&0&\\[1.5ex]
\text{\scriptsize{forward BC}}&\displaystyle n_\beta \cdot \overrightarrow{\nabla}_\beta E(\beta,t|x,s)&=&\kappa(\beta) E(\beta,t|x,s)& \beta \in \partial D^r\\[1.5ex]
\text{\scriptsize{backward BC}}&\displaystyle E(y,t|\beta,s) \overleftarrow{\nabla}_\beta \cdot n_\beta &=&\kappa(\beta) E(y,t|\beta,s)& \beta \in \partial D^r\\[1.5ex]
\text{\scriptsize{forward STC}}&\displaystyle\lim_{s \nearrow t} E(y,t|x,s)&=&\delta(|y-x|)&\\[1.5ex]
\text{\scriptsize{backward STC}}&\displaystyle\lim_{t \searrow s} E(y,t|x,s)&=&\delta(|y-x|)&\\
\end{array}
\end{equation}
where the BCs are satisfied because a Brownian particle is either reflected in the normal direction or absorbed with a certain probability. For each infinitesimal unit of time spent on the boundary, the probability of absorption is equal to $\kappa(\beta) dt$ and that of reflection $(1-\kappa(\beta))dt$, where $\kappa(\beta)\geq0$ and may depend on $\beta$. When $\kappa$ is constant and when a total time of $t$ is spent on the boundary, consisting of $n$ infinitesimal units $dt$, then the probability of survival becomes $(1-\kappa\,t/n)^n\to e^{-\kappa\,t}$.

\subsection{First- and last-passage (or reflection) decompositions}

The original research on Brownian motion starts here. We start by writing down the following identities for ABM:
\begin{equation}
\label{Identities}
\makebox[13cm][l]
{
$
\begin{array}{lr@{\hspace{1mm}}c@{\hspace{1mm}}l@{\hspace{1mm}}r@{\hspace{1mm}}c@{\hspace{1mm}}l}
\text{\tiny{FP}}&A(y,t|x,s)&=&B(y,t|x,s)&+&\displaystyle{\int _s^td\tau \,\frac{\partial}{\partial{\tau}}}&\displaystyle{\underset{D\hspace{2mm}}{\int } d\alpha\;B(y,t|\alpha,\tau)A(\alpha,\tau|x,s)}\\
\text{\tiny{LP}}&A(y,t|x,s)&=&B(y,t|x,s)&-&\displaystyle{\int _s^td\tau  \,\frac{\partial}{\partial{\tau}}} &\displaystyle{\underset{D\hspace{2mm}}{\int} d\alpha\;A(y,t|\alpha,\tau)B(\alpha,\tau|x,s)}
\end{array}
$
}
\end{equation}
where these indenties hold by the virtue of the fundamental theorem of calculus and the STCs satisfied by $B$ and $A$. First, by the fundamental theorem of calculus we have:
\begin{equation}
\makebox[13cm][l]
{
$
\begin{array}{lr@{\hspace{1mm}}c@{\hspace{1mm}}l@{\hspace{1mm}}r@{\hspace{1mm}}c@{\hspace{1mm}}l}
\text{\tiny{FP}}&A(y,t|x,s)&=&B(y,t|x,s)&+&\displaystyle{\left(\underset{\tau \nearrow t}\lim-\underset{\tau \searrow s}\lim\right)}&\displaystyle{\underset{D}{\int } d\alpha\;B(y,t|\alpha,\tau)A(\alpha,\tau|x,s)}\\
\text{\tiny{LP}}&A(y,t|x,s)&=&B(y,t|x,s)&-&\displaystyle{\left(\underset{\tau \nearrow t}\lim-\underset{\tau \searrow s}\lim\right)}&\displaystyle{\underset{D}{\int} d\alpha\;A(y,t|\alpha,\tau)B(\alpha,\tau|x,s)}
\end{array}
$
}
\end{equation}
and, second, the STCs satisfied by $B$ and $A$ show that both identities hold as long as $x$ and $y$ are in the interior of $D$. The nomenclature of first-passage (FP) and last-passage (LP) decompositions will now be explained. We define the first-passage time $\tau ^{\text{\tiny{FP}}}$ as the first time in the interval $[s,t]$ that $\partial D$ is crossed, or as infinite if there is no passage in $[s,t]$, i.e. $\inf\{\varnothing\}=\infty$. Similarly, we define the last-passage time $\tau ^{\text{\tiny{LP}}}$ as the last time in the interval $[s,t]$ that $\partial D$ is crossed, or as negative infinity if there is no passage in $[s,t]$, i.e. $\sup\{\varnothing\}=-\infty$. Therefore the quantity
\begin{equation*}
\underset{D}{\int} d\alpha\,B(y,t|\alpha,\tau)A(\alpha,\tau|x,s)
\end{equation*} 
counts paths from $(x,s)$ to $(y,t)$ where the first passage (if at all) happens \textit{after} time $\tau$. With these definitions, we can write:
\begin{equation}
\makebox[13cm][l]
{
$
\begin{array}{lr@{\hspace{1mm}}c@{\hspace{1mm}}l@{\hspace{1mm}}r@{\hspace{1mm}}c@{\hspace{1mm}}l}
\text{\tiny{FP}}&\mathbb{P}\left(B_t \in dy\text{; }\tau ^{\text{\tiny{FP}}}\geq\tau|B_s=x\right)&=&\displaystyle\underset{D}{\int} d\alpha\,  B(y,t|\alpha,\tau)A(\alpha,\tau|x,s),\\
\text{\tiny{LP}}&\mathbb{P}\left(B_t \in dy\text{; }\tau ^{\text{\tiny{LP}}}\leq\tau|B_s=x\right)&=&\displaystyle\underset{D}{\int} d\alpha\,  A(y,t|\alpha,\tau)B(\alpha,\tau|x,s),
\end{array}
$
}
\end{equation}
where the semi-colon indicates a joint probability. The free propagator $B$ allows passages while not requiring them, and therefore it is crucial that we specified that  
$\inf\{\varnothing\}=\infty$ and $\sup\{\varnothing\}=-\infty$. Differentiating, we get
\begin{equation}
\label{differentiated}
\makebox[13cm][l]
{
$
\begin{array}{lr@{\hspace{1mm}}c@{\hspace{1mm}}r@{\hspace{1mm}}l}
\text{\tiny{FP}}&\mathbb{P}\left(B_t \in dy\text{; }\tau ^{\text{\tiny{FP}}}\in d\tau|B_s=x\right)&=&\displaystyle-\frac{\partial}{\partial{\tau}}&\displaystyle\underset{D}{\int} d\alpha\,  B(y,t|\alpha,\tau)A(\alpha,\tau|,s)\\
\text{\tiny{LP}}&\mathbb{P}\left(B_t \in dy\text{; }\tau ^{\text{\tiny{LP}}}\in d\tau|B_s=x\right)&=&\displaystyle\frac{\partial}{\partial{\tau}}&\displaystyle\underset{D}{\int} d\alpha\,  A(y,t|\alpha,\tau)B(\alpha,\tau|,s)
\end{array}
$
}
\end{equation}
The absorbed density requires that no passages occur; first nor last passages. Therefore we subtract from the free density the integral (over $\tau$) of all paths with a first or last passage at time $\tau$, i.e.
\begin{equation}
\makebox[13cm][l]
{
$
\begin{array}{lr@{\hspace{1mm}}c@{\hspace{1mm}}l@{\hspace{1mm}}r@{\hspace{1mm}}l@{\hspace{1mm}}c@{\hspace{1mm}}l}
\text{\tiny{FP}}&A(y,t|x,s)&=&B(y,t|x,s)&+&\displaystyle\int _s^td\tau & \displaystyle \frac{\partial}{\partial{\tau}}&\displaystyle{\underset{D}{\int } d\alpha\;B(y,t|\alpha,\tau)A(\alpha,\tau|x,s)}\\
\text{\tiny{LP}}&A(y,t|x,s)&=&B(y,t|x,s)&-&\displaystyle\int _s^td\tau & \displaystyle \frac{\partial}{\partial{\tau}}&\displaystyle{\underset{D}{\int} d\alpha\;A(y,t|\alpha,\tau)B(\alpha,\tau|x,s)}
\end{array}
$
}
\end{equation}
and we have re-derived our identities, but now with a probabilistic intuition. In the `derivation' of these identities, we have used the STCs but not the PDEs or BCs. Differentiation under the integral sign is allowed and we can use the PDEs of \eqref{AbsorbedBrownianMotion}, to obtain
\begin{equation}
\makebox[13cm][l]
{
$
\begin{array}{lr@{\hspace{1mm}}c@{\hspace{1mm}}l}
\text{\tiny{FP}}&A(y,t|x,s)&=&B(y,t|x,s)-\displaystyle{\frac{\sigma ^2}{2}\int _s^td\tau \underset{D}{\int} d\alpha\;B(y,t|\alpha,\tau)\left \{\overleftarrow\nabla_\alpha^2-\overrightarrow\nabla_\alpha^2\right\}A(\alpha,\tau|x,s)}\\
\text{\tiny{LP}}&A(y,t|x,s)&=&B(y,t|x,s)+\displaystyle{\frac{\sigma ^2}{2}\int _s^td\tau  \underset{D}{\int} d\alpha\;A(y,t|\alpha,\tau)\left\{\overleftarrow\nabla_\alpha^2-\overrightarrow\nabla_\alpha^2\right\}B(\alpha,\tau|x,s)}
\end{array}
$
}
\end{equation}
where the direction of the arrows indicates the direction of differentiation. We feel that this notation makes equations more readable. Then we use Green's second identity --- which is valid for domains with a finite number of edges, corners and cusps --- to obtain
\begin{equation}
\label{AbsorbedGreenTheorem}
\makebox[13cm][l]
{
$
\begin{array}{lr@{\hspace{1mm}}c@{\hspace{1mm}}l}
\text{\tiny{FP}}&A(y,t|x,s)&=&B(y,t|x,s)+\displaystyle{\frac{1}{2}\int _s^td\tau \underset{\partial D}{\oint} d\beta\,\,B(y,t|\beta,\tau)\left\{\overleftarrow{\partial_\beta}-\overrightarrow{\partial_\beta}\right\}A(\beta,\tau|x,s)}\\
\text{\tiny{LP}}&A(y,t|x,s)&=&B(y,t|x,s)-\displaystyle{\frac{1}{2}\int _s^td\tau  \underset{\partial D}{\oint} d\beta\,\,A(y,t|\beta,\tau)\left\{\overleftarrow{\partial_\beta}-\overrightarrow{\partial_\beta}\right\}B(\beta,\tau|x,s)}\\
\end{array}
$
}
\end{equation}
where $\partial_\beta$ is the (by $\sigma^2$) scaled inward normal derivative, i.e. 
\begin{equation}
\label{DifferentialOperators}
\begin{array}{r@{\hspace{2mm}}l}
\overrightarrow{\partial_\beta} f(\beta,\gamma):=&- \sigma^2 {\underset{\alpha \rightarrow \beta }{\lim}}\, n_{\beta } \cdot \overrightarrow{\nabla}_{\alpha }f(\alpha,\gamma)\\
f(\gamma,\beta) \overleftarrow{\partial_\beta} :=&- \sigma^2 {\underset{\alpha \rightarrow \beta }{\lim}}\, n_{\beta } \cdot \overrightarrow{\nabla}_{\alpha }f(\gamma,\alpha)\\
\end{array}
\end{equation}
and where $\alpha$ is an interior coordinate, and where $\beta$ is a regular boundary coordinate. The BCs require that $A$ is zero on all regular parts of the boundary, and since the irregular parts have zero measure on the surface, we are left with the following:
\begin{equation}
\label{AbsorbedIntegralEquations}
\makebox[13cm][l]
{
$
\begin{array}{lr@{\hspace{1mm}}c@{\hspace{1mm}}l}
\text{\tiny{FP}}&A(y,t|x,s)&=&B(y,t|x,s)-\displaystyle{\int _s^t d\tau \underset{\partial D^r}{\oint} d\beta\;B(y,t|\beta,\tau)\left\{\frac{1}{2}\overrightarrow{\partial_\beta}\right\}A(\beta,\tau|x,s)}\\
\text{\tiny{LP}}&A(y,t|x,s)&=&B(y,t|x,s)-\displaystyle{\int _s^t d\tau  \underset{\partial D^r}{\oint} d\beta\;A(y,t|\beta,\tau)\left\{\frac{1}{2}\overleftarrow{\partial_\beta}\right\}B(\beta,\tau|x,s)}
\end{array}
$
}
\end{equation}
where the differential operators point towards the absorbed density $A$ in both cases. We also notice that a positive term is subtracted from the free density to obtain the absorbed density. In the last pair of equations, it should make no difference whether the integration is over the entire boundary $\partial D$ or over its regular part $\partial D^r$, since there is only a supposed to be a finite number of irregular boundary points.  

The FP decomposition has been obtained before, through the \textit{path decomposition expansion} that was developed in \cite{Auerbach1984} and \cite{Auerbach1985} and extended by \cite{Goodman1981} and \cite{Halliwell1995}. Both original and subsequent proofs rely on a detailed treatment involving time-slicing and  limit in which the number of slices goes to infinity. Our derivation is arguably simpler (fewer steps and a clear intuition), we derive double the result (i.e. two decompositions rather than one), we put an explicit requirement on the domain (i.e. Green's second identity), and allow a generalisation to other boundary conditions (for e.g. a reflecting boundary, see below). 

On a related but different note, we observe that the absorbed propagator $A$ is symmetric in the spatial coordinates $x$ and $y$, as a direct consequence of the FP/LP pair in \eqref{AbsorbedIntegralEquations}. This deserves some attention, since Chung, for example, writes \cite[p.~90]{Chung1995}):
\begin{quote}
By the way, there is NO probabilistic intuition for the symmetry of [the absorbed transition density].
\end{quote}
where the capitals appear in the reference. Our set of equations, however, can easily be interpreted when we realise that
\begin{equation}
\makebox[13cm][l]
{
$
\begin{array}{ll@{\hspace{1mm}}c@{\hspace{1mm}}l}
\text{\tiny{FP}}&\mathbb{P}\big({B_t\in  dy \text{; }\tau^\text{\tiny{FP}}}\in d\tau\text{; }B_{\tau^{\text{\tiny{FP}}}} \in d\beta  \big|B_s=x\big)&=&\displaystyle{B(y,t|\beta,\tau)\left\{\frac{1}{2}\overrightarrow{\partial_\beta}\right\} A(\beta,\tau|x,s)}\\[3ex]
\text{\tiny{LP}}&\mathbb{P}\big(B_t\in  dy \text{; }\tau^{\text{\tiny{LP}}}\in d\tau\text{; }B_{\tau ^{\text{\tiny{LP}}}}\in  d\beta\big|B_s=x\big)&=&\displaystyle{A(y,t|\beta ,\tau)\left\{\frac{1}{2}\overleftarrow{\partial_\beta}\right\} B(\beta,\tau|x,s)}
\end{array}
$
}
\end{equation}
And thus the spatial symmetry follows ultimately from a time reversal, where first becomes last and vice versa. Chung himself notes in \cite{Chung1973} why last-passage times are much less popular than first-passage times: the last passage is not a stopping time (i.e. it cannot be known immediately after the last passage that it is, indeed, the last passage). Chung argues that it is desirable that the first- and last-passage times are put on equal footing, and our approach does exactly that. 

We have used all six PDEs, STCs and BCs of \eqref{AbsorbedBrownianMotion} in the derivation of the pair \eqref{AbsorbedIntegralEquations}, i.e. all conditions that specify $A$ uniquely have now been used ---  along with Green's second identity on the domain. We state the following proposition:

\begin{absorbed} \label{Proposition1} \textnormal{\textbf{FP and LP decompositions of ABM.}} For all $x,y \in D$, where $D$ allows Green's theorem, and for all $\beta \in \partial D^r$, the following formulations of ABM are equivalent:
\begin{equation}
\left.
\begin{array}{r@{\hspace{1mm}}l}
\Big(\frac{\sigma^2}{2}\nabla_y^2-\frac{\partial}{\partial t}\Big)A(y,t|x,s)=&0\\[1.2ex]
\Big(\frac{\sigma^2}{2}\nabla_x^2+\frac{\partial}{\partial s}\Big)A(y,t|x,s)=&0\\[1.2ex]
A(\beta,t|x,s)=&0\\[1.2ex]
A(y,t|\beta,s)=&0\\[1.2ex]
\lim_{s \nearrow t} A(y,t|x,s)=&\delta(|y-x|)\\[1.2ex]
\lim_{t \searrow s} A(y,t|x,s)=&\delta(|y-x|)\\[1.2ex]
\end{array}
\right\}=\left\{
\begin{array}{ll}
\textnormal{\tiny{FP}}&A(y,t|x,s)=B(y,t|x,s)\\
&\;-\displaystyle{\int _s^td\tau \underset{\partial D^r}{\oint} d\beta\;B(y,t|\beta,\tau)\left\{\frac{1}{2}\overrightarrow{\partial_\beta}\right\}A(\beta,\tau|x,s)}\\[3ex]
\textnormal{\tiny{LP}}&A(y,t|x,s)=B(y,t|x,s)\\
&\;-\displaystyle{\int _s^td\tau  \underset{\partial D^r}{\oint} d\beta\;A(y,t|\beta,\tau)\left\{\frac{1}{2}\overleftarrow{\partial_\beta}\right\}B(\beta,\tau|x,s)}\\[.2ex]
\end{array}
\right.
\end{equation}
where $\partial_\beta$ is the (by $\sigma^2)$ scaled inward normal derivative as defined in \eqref{DifferentialOperators}, and where the arrow indicates the direction of the differentiation.
\end{absorbed}

Looking back at equation \eqref{AbsorbedGreenTheorem}, we could also have changed the signs of terms that are zero, instead of discarding them. This would lead to
\begin{equation}
\makebox[13cm][l]
{
$
\begin{array}{lr@{\hspace{1mm}}c@{\hspace{1mm}}l}
\text{\tiny{FP}}&A(y,t|x,s)&=&B(y,t|x,s)-\displaystyle{\frac{1}{2}\int _s^td\tau \underset{\partial D}{\oint} d\beta\,\,B(y,t|\beta,\tau)\left\{\overleftarrow{\partial_\beta}+\overrightarrow{\partial_\beta}\right\}A(\beta,\tau|x,s)}\\
\text{\tiny{LP}}&A(y,t|x,s)&=&B(y,t|x,s)-\displaystyle{\frac{1}{2}\int _s^td\tau  \underset{\partial D}{\oint} d\beta\,\,A(y,t|\beta,\tau)\left\{\overleftarrow{\partial_\beta}+\overrightarrow{\partial_\beta}\right\}B(\beta,\tau|x,s)}\\
\end{array}
$
}
\end{equation}
By the divergence theorem, we get
\begin{equation}
\makebox[13cm][l]
{
$
\begin{array}{lr@{\hspace{1mm}}c@{\hspace{1mm}}l}
\text{\tiny{FP}}&A(y,t|x,s)&=&B(y,t|x,s)+\displaystyle\frac{\sigma^2}{2}\int _s^td\tau \underset{D}{\int} d\alpha\,\,\nabla_\alpha^2 \bigg[B(y,t|\alpha,\tau)A(\alpha,\tau|x,s)\bigg]\\
\text{\tiny{LP}}&A(y,t|x,s)&=&B(y,t|x,s)+\displaystyle\frac{\sigma^2}{2}\int _s^td\tau  \underset{D}{\int} d\alpha\,\,\nabla_\alpha^2 \bigg[ A(y,t|\alpha,\tau)B(\alpha,\tau|x,s)\bigg]\\
\end{array}
$
}
\end{equation}
This does not have as clear a probabilistic intuition as Proposition \ref{Proposition1}, but these expressions are very suited to a series solution: substitute the definition of $A$, as given by the left-hand side, into the expression for $A$ on the right-hand side. The obtained series solution looks like the one obtained in e.g. \cite{Hsu1986}, by the \textit{parametrix method}. 

Having discussed the first- and last-passage decompositions at length, the following first-reflection (FR) and last-reflection (LR) decompositions suggest themselves:
\begin{equation}
\makebox[13cm][l]
{
$
\begin{array}{l@{\hspace{5mm}}l@{\hspace{1mm}}c@{\hspace{1mm}}l@{\hspace{1mm}}c@{\hspace{1mm}}l}
\text{\tiny{FR}}&R(y,t|x,s)&=&A(y,t|x,s)-\displaystyle\int _s^td\tau &\displaystyle\frac{\partial}{\partial\tau }&\displaystyle\underset{D}{\int} d\alpha\,R(y,t|\alpha,\tau)A(\alpha,\tau|x,s)\\[2ex]
\text{\tiny{LR}}&R(y,t|x,s)&=&A(y,t|x,s)+\displaystyle\int _s^td\tau &\displaystyle\frac{\partial}{\partial\tau} &\displaystyle\underset{D}{\int } d\alpha\, A(y,t|\alpha,\tau)R(\alpha,\tau|x,s)
\end{array}
$
}
\end{equation}
which hold by the virtue of the fundamental theorem of calculus and the STCs. It turns out, however, that it is more useful to write the reflected density in terms of the free density, and to do this we replace the absorbed density $A$ by the free density $B$:
\begin{equation}
\makebox[13cm][l]
{
$
\begin{array}{l@{\hspace{5mm}}l@{\hspace{1mm}}c@{\hspace{1mm}}l@{\hspace{1mm}}c@{\hspace{1mm}}l}
\text{\tiny{FR}}&R(y,t|x,s)&=&B(y,t|x,s)-\displaystyle\int _s^td\tau &\displaystyle\frac{\partial}{\partial\tau }&\displaystyle\underset{D}{\int} d\alpha\,R(y,t|\alpha,\tau)B(\alpha,\tau|x,s)\\[2ex]
\text{\tiny{LR}}&R(y,t|x,s)&=&B(y,t|x,s)+\displaystyle\int _s^td\tau &\displaystyle\frac{\partial}{\partial\tau}&\displaystyle\underset{D}{\int } d\alpha\, B(y,t|\alpha,\tau)R(\alpha,\tau|x,s)
\end{array}
$
}
\end{equation}
where we have kept the names FR and LR, even though that interpretation has now become a little bit problematic. But both identities still hold by the virtue of the fundamental theorem of calculus and the STCs, and proceeding with the same steps as in the absorbed case, we obtain:

\begin{reflected} \label{Proposition2} \textnormal{\textbf{FR and LR decompositions of RBM.}} 
For all $x,y \in D$, where $D$ allows Green's theorem, and for all $\beta \in \partial D^r$, the following formulations of RBM are equivalent:
\begin{equation}
\left.
\begin{array}{r@{\hspace{1mm}}l}
\Big(\frac{\sigma^2}{2}\nabla_y^2-\frac{\partial}{\partial t}\Big)R(y,t|x,s)=&0\\[1.2ex]
\Big(\frac{\sigma^2}{2}\nabla_x^2+\frac{\partial}{\partial s}\Big)R(y,t|x,s)=&0\\[1.2ex]
\overrightarrow{\partial_\beta} R(\beta,t|x,s)=&0\\[1.2ex]
R(y,t|\beta,s)\overleftarrow{\partial_\beta}=&0\\[1.2ex]
\lim_{s \nearrow t} R(y,t|x,s)=&\delta(|y-x|)\\[1.2ex]
\lim_{t \searrow s} R(y,t|x,s)=&\delta(|y-x|)\\
\end{array}
\right\}=\left\{
\begin{array}{ll}
\textnormal{\tiny{FR}}&R(y,t|x,s)=B(y,t|x,s)\\
&\;+\displaystyle{\int _s^td\tau \underset{\partial D^r}{\oint} d\beta\;R(y,t|\beta,\tau)\left\{\frac{1}{2}\overrightarrow{\partial_\beta}\right\}B(\beta,\tau|x,s)}\\[4ex]
\textnormal{\tiny{LR}}&R(y,t|x,s)=B(y,t|x,s)\\
&\;+\displaystyle{\int _s^td\tau  \underset{\partial D^r}{\oint} d\beta\;B(y,t|\beta,\tau)\left\{\frac{1}{2}\overleftarrow{\partial_\beta}\right\}R(\beta,\tau|x,s)}\end{array}
\right.
\end{equation}
where $\partial_\beta$ is the (by $\sigma^2)$ scaled inward normal derivative as defined in \eqref{DifferentialOperators}, and where the arrow indicates the direction of the differentiation.
\end{reflected}
Whereas the absorbed density $A$ is always smaller than the free density $B$, it is \textit{not} the case that the reflected density is always larger than the free density $B$. The reflected density equals the free density $B$ plus a weighted average of boundary densities of $R$. We have that
\begin{equation*}
\left\{\frac{1}{2}\overrightarrow{\partial_\beta}\right\}B(\beta,\tau|x,s)=n_\beta \cdot \frac{\beta-x}{t-s} B(\beta,t|x,s)
\mbox{ $\geq 0$ if $D$ is convex}\\
\end{equation*}
with strict inequalities if $D$ is strictly convex. Thus for a convex space, the reflected density is everywhere larger than the free density. Intuitively, every point in a convex domain is like a `focal' point, where more paths are directed than in the absence of the boundary.

The same exercise can be repeated for elastic Brownian motion, to obtain that \eqref{ElasticBrownianMotion} is equal to 
\begin{equation}
\makebox[13cm][l]
{
$
\begin{array}{ll}
\textnormal{\tiny{FR}}&E(y,t|x,s)=B(y,t|x,s)+\displaystyle{\int _s^td\tau \underset{\partial D^r}{\oint} d\beta\;E(y,t|\beta,\tau)\left\{\frac{1}{2}\overrightarrow{\partial_\beta}-\frac{\sigma^2}{2}\kappa(\beta)\right\}B(\beta,\tau|x,s)}\\[4ex]
\textnormal{\tiny{LR}}&E(y,t|x,s)=B(y,t|x,s)+\displaystyle{\int _s^td\tau  \underset{\partial D^r}{\oint} d\beta\;B(y,t|\beta,\tau)\left\{\frac{1}{2}\overleftarrow{\partial_\beta}-\frac{\sigma^2}{2}\kappa(\beta)\right\}E(\beta,\tau|x,s)}
\end{array}
$
}
\end{equation}
In the limit where $\kappa(\beta)\searrow 0$, we recover the integral equations governing RBM.

\subsection{Tangent-plane decompositions and series solution}

The integral equations of Proposition \ref{Proposition1} have the property that the absorbed density $A$ appears on the left-hand side, while the normal derivative of $A$ appears on the right-hand side. The trick for solving integral equations like this is to make sure that the \textit{same} quantity appears on both sides of the equation. Therefore we would like the normal derivative to appear on both sides. Thus we apply $\frac{1}{2}\overrightarrow{\partial_\beta}$ to the left of the FP decomposition, and $\frac{1}{2}\overleftarrow{\partial_\beta}$ to the right of the LP decomposition, i.e.
\begin{equation}
\makebox[13cm][l]
{
$
\begin{array}{lr@{\hspace{1mm}}c@{\hspace{1mm}}l}
\text{\tiny{FP}}&\displaystyle\left\{\frac{1}{2}\overrightarrow{\partial_\beta}\right\}A(\beta,t|x,s)&=&\displaystyle\left\{\frac{1}{2}\overrightarrow{\partial_\beta}\right\}B(\beta,t|x,s)\\
&&&-\displaystyle\left\{\frac{1}{2}\overrightarrow{\partial_\beta}\right\}\int _s^t d\tau \underset{\partial D}{\oint} d\gamma\;B(\beta,t|\gamma,\tau)\left\{\frac{1}{2}\overrightarrow{\partial_\gamma}\right\}A(\gamma,\tau|x,s)\\
\text{\tiny{LP}}&\displaystyle A(y,t|\beta,s)\left\{\frac{1}{2}\overleftarrow{\partial_\beta}\right\}&=&\displaystyle B(y,t|\beta,s)\left\{\frac{1}{2}\overleftarrow{\partial_\beta}\right\}\\
&&&-\displaystyle\left(\int _s^t d\tau  \underset{\partial D}{\oint} d\gamma\;A(y,t|\gamma,\tau)\left\{\frac{1}{2}\overleftarrow{\partial_\gamma}\right\}B(\gamma,\tau|\beta,s)\right)\left\{\frac{1}{2}\overleftarrow{\partial_\beta}\right\}
\end{array}
$
}
\end{equation}
Using Lemma \ref{Lemma1} in section \ref{Lemmas} to push the differential boundary operators through the integral signs, we get
\begin{equation}
\makebox[13cm][l]
{
$
\begin{array}{lr@{\hspace{1mm}}c@{\hspace{1mm}}l}
\text{\tiny{FP}}&\displaystyle\left\{\frac{1}{2}\overrightarrow{\partial_\beta}\right\}A(\beta,t|x,s)&=&\displaystyle\left\{\frac{1}{2}\overrightarrow{\partial_\beta}\right\}B(\beta,t|x,s)\\
&&&-\displaystyle \int _s^t d\tau \underset{\partial D}{\oint} d\beta\;\left\{\frac{1}{2}\overrightarrow{\partial_\beta}\right\}B(\beta,t|\gamma,\tau)\left\{\frac{1}{2}\overrightarrow{\partial_\gamma}\right\}A(\gamma,\tau|x,s)\\
&&&+\displaystyle \frac{1}{2} \left\{\frac{1}{2}\overrightarrow{\partial_\beta}\right\}A(\beta,t|x,s)\\
\text{\tiny{LP}}&\displaystyle A(y,t|\beta,s)\left\{\frac{1}{2}\overleftarrow{\partial_\beta}\right\}&=&\displaystyle B(y,t|\beta,s)\left\{\frac{1}{2}\overleftarrow{\partial_\beta}\right\}\\
&&&-\displaystyle \int _s^t d\tau  \underset{\partial D}{\oint} d\gamma\;A(y,t|\gamma,\tau)\left\{\frac{1}{2}\overleftarrow{\partial_\gamma}\right\}B(\gamma,\tau|\beta,s)\left\{\frac{1}{2}\overleftarrow{\partial_\beta}\right\}\\
&&&+\displaystyle \frac{1}{2} A(y,t|\beta,s) \left\{\frac{1}{2}\overleftarrow{\partial_\beta}\right\}\\
\end{array}
$
}
\end{equation}
Collecting terms, we obtain what we shall call the \textit{tangent-plane decomposition}:
\begin{equation}
\label{AbsorbedTangentPlane}
\makebox[13cm][l]
{
$
\begin{array}{lr@{\hspace{1mm}}c@{\hspace{1mm}}l}
\text{\tiny{FP}}&\displaystyle\left\{\frac{1}{2}\overrightarrow{\partial_\beta}\right\}A(\beta,t|x,s)&=&\displaystyle \overrightarrow{\partial_\beta}B(\beta,t|x,s)\\
&&&-\displaystyle \int _s^t d\tau \underset{\partial D}{\oint} d\beta\; \overrightarrow{\partial_\beta} B(\beta,t|\gamma,\tau)\left\{\frac{1}{2}\overrightarrow{\partial_\gamma}\right\}A(\gamma,\tau|x,s)\\
\text{\tiny{LP}}&\displaystyle A(y,t|\beta,s)\left\{\frac{1}{2}\overleftarrow{\partial_\beta}\right\}&=&\displaystyle B(y,t|\beta,s)\overleftarrow{\partial_\beta}\\
&&&-\displaystyle \int _s^t d\tau  \underset{\partial D}{\oint} d\gamma\;A(y,t|\gamma,\tau)\left\{\frac{1}{2}\overleftarrow{\partial_\gamma}\right\}B(\gamma,\tau|\beta,s)\overleftarrow{\partial_\beta}\\
\end{array}
$
}
\end{equation}	
where the factors of 2 are crucial and the factorisation is carefully chosen. Fist, we notice from the first-passage decomposition that the first-passage density at $(\beta,t)$, which appears on the left-hand side, is related to the first-passage density at all other space-time locations $(\gamma,\tau)$, for all $\gamma$ and $\tau$. This was to be expected, since the shape of the entire boundary influences the first-passage density at any single location. 

The only case when the first-passage density decouples from other locations is when the domain is halfspace: the first-passage density for a halfspace consists only of the first term in \eqref{AbsorbedTangentPlane}. This is because the second term in \eqref{AbsorbedTangentPlane} equals zero, i.e.  
\begin{equation*}
\int_s^t d\tau \underset{\partial D}{\oint} d\gamma\;\overrightarrow{\partial_\beta} B(\beta,t|\gamma,\tau)\left\{\frac{1}{2}\overrightarrow{\partial_\gamma}\right\} A(\gamma,\tau|x,s)
\end{equation*}
equals zero for a halfspace, because 
\begin{equation*}
\overrightarrow{\partial_\beta} B(\beta,t|\gamma,\tau)=n_\beta \cdot \frac{\beta-\gamma}{t-\tau} B(\beta,t|\gamma,\tau)
\end{equation*} 
where for a halfspace we have that $ n_\beta \cdot (\beta-\gamma)=0$ because $n_\beta$ and $(\beta-\gamma)$ are perpendicular. For a halfspace, therefore, the first term in the tangent-plane decomposition is the only term. 

In general, we conclude that the first-passage density at any location $\beta$, depends on the first-passage density at all other locations $\gamma$ through a certain `weight', where this weight can be positive or negative and takes the sign of $n_\beta \cdot (\beta-\gamma)$. It is not hard to check that the following variational inequalities hold for convex and concave spaces:
\begin{equation}
\begin{array}{r@{\hspace{6mm}}c@{\hspace{6mm}}l}
\text{Convex domain}&n_\beta \cdot (\beta-\gamma)\geq 0&\mbox{$\beta,\gamma \in \partial D$}\\
\text{Convave domain}&n_\beta \cdot (\beta-\gamma)\leq 0&\mbox{$\beta,\gamma \in \partial D$}
\end{array}
\end{equation}
As a result, the first-passage density at any location of a convex domain is \textit{smaller} than the corresponding first-passage density over the tangent plane at that location. And the opposite holds for a concave domain. 

Consider, for the time being, a convex domain $D$. Then $D$ has the property that \textit{all} the tangent planes lie outside of $D$. If a particle is not allowed to leave $D$, then it is not allowed to cross any of the tangent planes defined by $\partial D$. The joint probability that a first-passage occurs at at the space-time coordinate $(\beta,t)$ can be estimated by the probability that the particle hits the tangent plane defined by $\beta$ for the first time at $(\beta,t)$ --- but this is an overestimate. Therefore we must subtract from this initial estimate the probability that the particle leaves the domain at some other space-time location $(\gamma,\tau)$ and \textit{then} hits the tangent plane defined by $\beta$ at $(\beta,t)$ ---  and we should sum over all $\gamma$ and $\tau$. We see that the right-hand side of the FP decomposition in \eqref{AbsorbedTangentPlane} does exactly this. This interpretation is new. 

The tangent-plane decompositions are useful not only because of their interpretation, but also because they feature the \textit{same} quantity on both sides of the equation. The idea for solving integral equations like \eqref{AbsorbedTangentPlane} is by the `successive approximation method' as in \cite[p. 566, 632, 811]{Polyanin1998} or, equivalently, by the `Neumann series' as in \cite[p.~78]{Porter1990}. The idea is simple: use the left-hand side of the equation as the definition for the unknown quantity appearing on the right-hand side, and do this repeatedly to obtain an infinite series solution. Once a series solution for $\partial A$  has been obtained, we can substitute it back into the expression for $A$ itself. If the series converges then it must be the answer that we are looking for, and thus we conclude:

\begin{AbsorbedSeries} \label{Proposition3} \textnormal{\textbf{Formal ABM series solution}}. The formal solution to problem \eqref{AbsorbedBrownianMotion} is given by the following first- or last-passage series:
\begin{equation*}
\makebox[13cm][l]
{
$
\begin{array}{lr@{\hspace{1mm}}c@{\hspace{1mm}}l}
\textnormal{\tiny{FP}}&A(y,t|x,s)&=&B(y,t|x,s)+\displaystyle\sum_{i=1}^{\infty}(-1)^i \Bigg[\underset{s\leq \theta_1\leq \ldots \leq \theta_i \leq t}{\int d\theta_i\ldots\int d\theta_1}\Bigg]\;\Bigg[\oint d\beta_i\ldots\oint d\beta_1\Bigg]\\
&&&\displaystyle \times B(y,t|\beta_i,\theta_i)\Bigg[\prod_{k=2}^{i} \overrightarrow{\partial_{\beta_{k}}}B(\beta_{k},\theta_k|\beta_{k-1},\theta_{k-1})\Bigg] \overrightarrow{\partial_{\beta_1}} B(\beta_1,\theta_1|x,s)\\[5ex]
\textnormal{\tiny{LP}}&A(y,t|x,s)&=&B(y,t|x,s)+\displaystyle\sum_{i=1}^{\infty}(-1)^i \Bigg[\underset{s\leq \theta_1\leq \ldots \leq \theta_i \leq t}{\int d\theta_i\ldots\int d\theta_1}\Bigg]\;\Bigg[\oint d\beta_i\ldots\oint d\beta_1\Bigg]\\
&&&\displaystyle \times B(y,t|\beta_i,\theta_i)\overleftarrow{\partial_{\beta_i}} \Bigg[ \prod_{k=1}^{i-1} B(\beta_{k+1},\theta_{k+1}|\beta_k,\theta_k) \overleftarrow{\partial_{\beta_k}}\Bigg] B(\beta_1,\theta_1|x,s)\\
\end{array}
$
}
\end{equation*}
where the \textnormal{\footnotesize{FP}} and \textnormal{\footnotesize{LP}} series are identical, term-by-term, and where the modes of convergence are as follows:
\begin{center}
\begin{tabular}
{|l|c|r|}
\hline
\text{domain} &\text{mode of convergence} \\\hline
\textnormal{convex domain} & \textnormal{alternating}\\ \hline
\textnormal{concave domain} & \textnormal{monotone} \\
\hline
\end{tabular}\end{center}
but where convergence itself is taken for granted. 
\end{AbsorbedSeries}
For a halfspace, only the free term and the first perturbation term are non-zero. We obtain
\begin{equation*}
\begin{array}{r@{\hspace{1mm}}c@{\hspace{1mm}}l}
A^{\text{\tiny HS}}(y,t|x,s)&=&B(y,t|x,s)-\displaystyle{\int_s^t d\theta_1}\oint d\beta_1\,B(y,t|\beta_1,\theta_1)\overrightarrow{\partial_{\beta_1}} B(\beta_1,\theta_1|x,s)\\[2ex]
&=&B(y,t|x,s)-\displaystyle{\int_s^t d\theta_1}\oint d\beta_1\,B(y,t|\beta_1,\theta_1)\overrightarrow{\partial_{\beta_1}} B(\beta_1,\theta_1|x^*,s)\\[2ex]
&=&\displaystyle B(y,t|x,s)-B(y,t|x^*,s)
\end{array}
\end{equation*}
where $x^*$ is the mirror-coordinate of $x$ ($x^*=-x$ for $d=1$) and where we have reproduced the solution for a halfspace that is normally obtained through the method of images. In general, all terms survive and all terms in the \textit{square brackets} in Proposition \ref{Proposition3} are positive (negative) for a convex (concave) domain, and therefore the mode of convergence is alternating (depends on $x$). Unfortunately, the sign of $\overrightarrow{\partial_\beta} B(\beta,\theta|x,s)$ may change as $\beta$ moves along a concave boundary. Where it has a fixed sign, the series converges in a monotone fashion. Where it has a fixed but different sign, the series also converges in a monotone fashion, except in the other direction. Because we can split the series solution into two parts where both converge in a monotone fashion (albeit in other directions), we say simply that the series converges in a monotone fashion.

For the reflected density, we have found the following integral equations:
\begin{equation}
\label{ReflectionDecompositions}
\makebox[13cm][l]
{
$
\begin{array}{ll@{\hspace{1mm}}c@{\hspace{1mm}}l}
\text{\tiny{FR}}&R(y,t|x,s)&=&B(y,t|x,s)+\displaystyle\int _s^td\tau \underset{\partial D}{\oint} d\beta\; R(y,t|\beta,\tau)\left\{\frac{1}{2}\overrightarrow{\partial_\beta}\right\}B(\beta,\tau|x,s)\\[2ex]
\text{\tiny{LR}}&R(y,t|x,s)&=&B(y,t|x,s)+\displaystyle\int _s^td\tau \underset{\partial D}{\oint } d\beta\; B(y,t|\beta,\tau)\left\{\frac{1}{2}\overleftarrow{\partial_\beta}\right\}R(\beta,\tau|x,s)
\end{array}
$
}
\end{equation}
We apply the operators $\lim_{x \to \beta}$ and $\lim_{y \to \beta}$ and use Lemma  \ref{Lemma2} in \ref{Lemmas} to push the limits through the integrals, collect terms and obtain the reflected tangent-plane decompositions:
\begin{equation}
\makebox[13cm][l]
{
$
\begin{array}{ll@{\hspace{1mm}}c@{\hspace{1mm}}l}
\text{\tiny{FR}}&R(y,t|\beta,s)&=&2 B(y,t|\beta,s)+\displaystyle\int _s^t d\tau \underset{\partial D}{\oint} d\gamma\; R(y,t|\gamma,\tau)\overrightarrow{\partial_\gamma}B(\gamma,\tau|\beta,s)\\[2ex]
\text{\tiny{LR}}&R(\beta,t|x,s)&=&2 B(\beta,t|x,s)+\displaystyle\int _s^td\tau \underset{\partial D}{\oint } d\gamma\; B(\beta,t|\gamma,\tau)\overleftarrow{\partial_\gamma}R(\gamma,\tau|x,s)
\end{array}
$
}
\end{equation}
where the factors of 2 are crucial. As before, for a halfspace only the first term in the tangent-plane decomposition survives. For e.g. a convex domain, we may estimate the reflected probability density $R$ at $\beta$ as if there was (only) a reflecting tangent plane at $\beta$. This gives rise to the first term on the right-hand side, which is $2B$. But for a convex domain with a reflecting boundary, every location is like a focal point: more paths are directed there. Therefore we must add to the initial estimate the probability that the particle reflects off the boundary somewhere else, and only \textit{then} reaches the tangent plane at $\beta$ for the first time at $(\beta,t)$. By a repeated substitution we find that
\begin{ReflectedSeries}\label{Proposition4} \textnormal{\textbf{Formal RBM series solution.}} The formal solution to problem \eqref{ReflectedBrownianMotion} is given by the following first- and last-reflection series:
\begin{equation}
\makebox[13cm][l]
{
$
\begin{array}{lr@{\hspace{1mm}}c@{\hspace{1mm}}l}
\textnormal{\tiny{FR}}&R(y,t|x,s)&=&B(y,t|x,s)+\displaystyle\sum_{i=1}^{\infty}\Bigg[\underset{s\leq \theta_1\leq \ldots \leq \theta_i \leq t}{\int d\theta_i\ldots\int d\theta_1}\Bigg]\;\Bigg[\oint d\beta_i\ldots\oint d\beta_1\Bigg]\\
&&&\displaystyle \times B(y,t|\beta_i,\theta_i)\Bigg[\prod_{k=2}^{i} \overrightarrow{\partial_{\beta_{k}}}B(\beta_{k},\theta_k|\beta_{k-1},\theta_{k-1})\Bigg] \overrightarrow{\partial_{\beta_1}} B(\beta_1,\theta_1|x,s)\\[5ex]
\textnormal{\tiny{LR}}&R(y,t|x,s)&=&B(y,t|x,s)+\displaystyle\sum_{i=1}^{\infty}\Bigg[\underset{s\leq \theta_1\leq \ldots \leq \theta_i \leq t}{\int d\theta_i\ldots\int d\theta_1}\Bigg]\;\Bigg[\oint d\beta_i\ldots\oint d\beta_1\Bigg]\\
&&&\displaystyle \times B(y,t|\beta_i,\theta_i)\overleftarrow{\partial_{\beta_i}} \Bigg[ \prod_{k=1}^{i-1} B(\beta_{k+1},\theta_{k+1}|\beta_k,\theta_k) \overleftarrow{\partial_{\beta_k}}\Bigg] B(\beta_1,\theta_1|x,s)\\
\end{array}
$
}
\end{equation}
where the \textnormal{\footnotesize{FR}} and \textnormal{\footnotesize{LR}} series are identical, term-by-term, and where the mode of convergence is as follows:
\begin{center}
\begin{tabular}
{|l|c|r|}
\hline
\text{domain} &\text{mode of convergence} \\\hline
\textnormal{convex domain} & \textnormal{monotone}\\ \hline
\textnormal{concave domain} & \textnormal{alternating} \\
\hline
\end{tabular}\end{center}
but where convergence itself is taken for granted. 
\end{ReflectedSeries}
\noindent The only difference with the absorbed series is that all terms here appear with a positive sign. For a halfspace, only the free term and the first perturbation term are non-zero. We obtain
\begin{equation*}
\begin{array}{r@{\hspace{1mm}}c@{\hspace{1mm}}l}
R^{\text{\tiny HS}}(y,t|x,s)&=&B(y,t|x,s)+\displaystyle{\int_s^t d\theta_1}\oint d\beta_1\,B(y,t|\beta_1,\theta_1)\overrightarrow{\partial_{\beta_1}} B(\beta_1,\theta_1|x,s)\\[2ex]
&=&\displaystyle B(y,t|x,s)+B(y,t|x^{*},s)
\end{array}
\end{equation*}
where $x^*$ is the mirror-coordinate of $x$ ($x^*=-x$ for $d=1$). In general, are perturbation terms survive. Similar series for reflected Brownian motion have been derived by the \textit{parametrix method} in e.g. \cite[p. 261]{Hsu1986}, \cite{Hsu1985,Hsu1992}. The series solution is based on an \textit{ansatz}, only one of the two series is derived, and it is thought to hold for smooth domains only. 

For the elastic case, we obtain similarly
\begin{equation}
\label{elasticseries}
\hspace{-2cm}
\makebox[13cm][l]
{
$
\begin{array}{lr@{\hspace{1mm}}c@{\hspace{1mm}}l}
\textnormal{\tiny{FR}}&E(y,t|x,s)&=&B(y,t|x,s)+\displaystyle\sum_{i=1}^{\infty}\Bigg[\underset{s\leq \theta_1\leq \ldots \leq \theta_i \leq t}{\int d\theta_i\ldots\int d\theta_1}\Bigg]\;\Bigg[\oint d\beta_i\ldots\oint d\beta_1\Bigg]\\
&&&\hspace{-1cm}\displaystyle \times B(y,t|\beta_i,\theta_i)\Bigg[\prod_{k=2}^{i} \left\{\overrightarrow{\partial_{\beta_{k}}}-\kappa(\beta_k)\sigma^2\right\}B(\beta_{k},\theta_k|\beta_{k-1},\theta_{k-1})\Bigg] \left\{\overrightarrow{\partial_{\beta_1}}-\kappa(\beta_1)\sigma^2\right\} B(\beta_1,\theta_1|x,s)\\[5ex]
\textnormal{\tiny{LR}}&E(y,t|x,s)&=&B(y,t|x,s)+\displaystyle\sum_{i=1}^{\infty}\Bigg[\underset{s\leq \theta_1\leq \ldots \leq \theta_i \leq t}{\int d\theta_i\ldots\int d\theta_1}\Bigg]\;\Bigg[\oint d\beta_i\ldots\oint d\beta_1\Bigg]\\
&&&\hspace{-1cm}\displaystyle \times B(y,t|\beta_i,\theta_i)\left\{\overleftarrow{\partial_{\beta_i}}-\kappa(\beta_i)\sigma^2\right\} \Bigg[ \prod_{k=1}^{i-1} B(\beta_{k+1},\theta_{k+1}|\beta_k,\theta_k) \left\{\overleftarrow{\partial_{\beta_k}}-\kappa(\beta_k)\sigma^2\right\}\Bigg] B(\beta_1,\theta_1|x,s)\\
\end{array}
$
}
\end{equation}
It is obvious that this series solution becomes less useful if we let $\kappa \to \infty$, because the series then no longer converges. However, we do know that $E \to A$ as $\kappa \to \infty$, and we have a series solution for $A$. In practice, therefore, the situation is only problematic when $\kappa$ is quite large but not infinite. 

Berry and Dennis (\cite{Berry2008}) treat a $2d$ circle, where $\kappa(\phi)=\sin(\phi)/(1- \epsilon \cos(\phi))$ for $-\pi\leq \phi \leq \pi$. In the limit where $\epsilon \nearrow 1$ they obtain $\kappa(\phi)=\cot(\phi/2)$. This means that the boundary destroys particles at rate $|\kappa(\phi)|$ for $y>0$, while creating particles at rate $|\kappa(\phi)|$ for $y<0$. If $\kappa(\beta)d\beta$ is absolutely integrable, then the series solution \eqref{elasticseries} exists. But for $\epsilon \nearrow 1$, we obtain $\kappa(0+)=\infty$, $\kappa(0-)=-\infty$ in a non-integrable way. Such particle creation at an infinite rate is problematic since no equilibrium state will exist. Suppose a circle-boundary is fully absorbing for $y>0$, and reflecting/creating at rate $|\kappa(\beta)|$ for $y<0$, then some equilibrium distribution may exist, \textit{unless} the creation goes to infinity so quickly that more particles are created than can be absorbed. A particle can be absorbed only once, but it can visit the reflecting/creating boundary more than once, suggesting that this side will `dominate'. This view may complement the one presented by Marletta and Rozenblum \cite{Marletta2009}, who have commented on this issue from a  mathematical point of view. 

Lastly, we need to prove that the first- and last-passage (or reflection) series are identical, term by term. In short-hand, we first note that by Lemma \ref{Lemma3} (in section \ref{Lemmas}) we obtain
\begin{equation*}
\int\oint B\overrightarrow{\partial} B =\int\oint B\overleftarrow{\partial} B \mbox{ for $x,y\in D$}
\end{equation*}
which proves that the first perturbation terms are equal. Is it also the case that\begin{equation*}
\int\oint\int\oint B \overrightarrow{\partial} B \overrightarrow{\partial} B\overset{?}=\int\oint\int\oint B \overleftarrow{\partial} B \overleftarrow{\partial}B
\end{equation*}
Using Lemma \ref{Lemma3} in section \ref{Lemmas} twice, we obtain 
\begin{equation*}
\begin{array}{rcl}
\displaystyle\int\oint\int\oint B \overrightarrow{\partial} B \overrightarrow{\partial} B&=&\displaystyle
\int\oint\int\oint B \overleftarrow{\partial} B \overrightarrow{\partial} B+ \int\oint B \overrightarrow{\partial}B\\
&=&\displaystyle
\int\oint\int\oint B \overleftarrow{\partial} B \overleftarrow{\partial} B- \int\oint B \overleftarrow{\partial}B+\int\oint B \overrightarrow{\partial}B\\
&=&\displaystyle
\int\oint\int\oint B \overleftarrow{\partial} B \overleftarrow{\partial} B\\
\end{array}
\end{equation*}
where we must be careful not to obtain terms like
\begin{equation*}
\int\oint\int\oint B \overrightarrow{\partial}B\overleftarrow{\partial}B= \infty
\end{equation*}
and similarly for higher order perturbation terms.

\subsection{Green functions and single and double boundary layers}

Now suppose that a source at $x$ emits one Brownian particle, and we ask what the expected amount of time is that the particle spends in the neighbourhood of any location in space, given that we observe the Brownian particle for an (infinitely) long time. We have the free Green function $G_B$ as follows
\begin{equation}
G_B(y,x):=\mathbb{E}_x\int _s^{\infty} \delta(|B_\tau - y|) d\tau =\int_s^{\infty}B(y,\tau| x,s)\,d\tau
\end{equation}
\noindent For $d\geq3$ the free Green function is finite and the integration can be performed to give:
\begin{equation}
\label{FreeGreenFunctionExplicit}
G_B(y,x) =\frac{1}{\sigma^2}\frac{\Gamma(d/2-1)}{2 \pi ^{d/2}}|y-x| ^{2-d}
\end{equation}
\noindent where $\Gamma$ denotes the gamma function. Similarly, the Green function associated with ABM is defined by
\begin{equation}
G_A(y,x):=\int_{s}^\infty A(y,t|x,s)\,dt
\end{equation} 
and satisfies
\begin{equation}
\begin{array}{c}
\displaystyle\frac{\sigma^2}{2}\nabla^2_y G_A(y,x)=\displaystyle\frac{\sigma^2}{2}\nabla^2_x G_A(y,x)=-\delta(|y-x|)\\
G_A(\beta,x)=G_A(y,\beta)=0
\end{array}
\end{equation}
for all $x$ and $y$ in the interior and for all regular boundary coordinates $\beta$. In the Brownian interpretation, the absorbed Green function is zero at the boundary because an absorbed Brownian motion will spend zero time there. 

By integrating the series solutions of Proposition \ref{Proposition3} over time, we get two series for the absorbed Green function:
\begin{equation}
\begin{array}{lr@{\hspace{1mm}}c@{\hspace{1mm}}l}
\text{\tiny{FP}}&G_A(y,x)&=&G_B(y,x)\\
&&&+\displaystyle\sum_{i=1}^{\infty}(-1)^i \left[\oint d\beta_i\ldots\oint d\beta_1\right]\,G_B(y,\beta_i)\left[\prod_{k=2}^{i} \overrightarrow{\partial_{\beta_{k}}}G_B(\beta_{k},\beta_{k-1})\right] \overrightarrow{\partial_{\beta_1}} G_B(\beta_1,x)\\[3ex]
\text{\tiny{LP}}&G_A(y,x)&=&G_B(y,x)\\
&&&+\displaystyle\sum_{i=1}^{\infty}(-1)^i \left[\oint d\beta_i\ldots\oint d\beta_1\right]\,
G_B(y,\beta_i)\overleftarrow{\partial_{\beta_i}} \left[ \prod_{k=1}^{i-1} G_B(\beta_{k+1},\beta_k) \overleftarrow{\partial_{\beta_k}}\right] G_B(\beta_1,x)
\end{array}
\end{equation}
with the same modes of convergence as those in Proposition \ref{Proposition3}. To contrast our result with the literature, we provide the following Corollary:
\begin{CorollaryAbsorbed}
\textnormal{\textbf{$\bm{G_A}$ as SBL or DBL.}}
The absorbed Green function $G_A$ can be found by a double boundary layer \textnormal{(DBL)} or single boundary layer \textnormal{(SBL)}:
\begin{equation}
\begin{array}{lr@{\hspace{1mm}}c@{\hspace{1mm}}l}
\textnormal{\tiny{FP}}&G_A(y,x)&=&\displaystyle G_B(y,x) -\int_{\partial D} d\beta\, \mu_{\textnormal{\tiny{DBL}}}(y,\beta) \overrightarrow{\partial_\beta} G_B(\beta,x)\\
\textnormal{\tiny{LP}}&G_A(y,x)&=&\displaystyle G_B(y,x)-\int_{\partial D} d\beta\, \mu_\textnormal{\tiny{SBL}}(y,\beta) G_B(\beta,x)\\
\end{array}
\end{equation}
with the following definitions of $\mu_{\textnormal{\tiny{DBL}}}$ and $\mu_\textnormal{\tiny{SBL}}$:
\begin{equation}
\begin{array}{lr@{\hspace{1mm}}c@{\hspace{1mm}}l}
\textnormal{\tiny{FP}}&\mu_{\textnormal{\tiny{DBL}}}(y,\beta)&=&G_B(y,\beta)\\
&&&+\displaystyle\sum_{i=1}^{\infty}(-1)^i \left[\oint d\beta_i\ldots\oint d\beta_1\right]\,G_B(y,\beta_i)\left[\prod_{k=2}^{i} \overrightarrow{\partial_{\beta_{k}}}G_B(\beta_{k},\beta_{k-1})\right] \overrightarrow{\partial_{\beta_1}} G_B(\beta_1,\beta)\\
\textnormal{\tiny{LP}}&\mu_\textnormal{\tiny{SBL}}(y,\beta)&=&G_B(y,\beta)\overleftarrow{\partial_\beta}\\
&&&+\displaystyle\sum_{i=1}^{\infty}(-1)^i \left[\oint d\beta_i\ldots\oint d\beta_1\right]\,
G_B(y,\beta_i)\overleftarrow{\partial_{\beta_i}} \left[ \prod_{k=1}^{i-1} G_B(\beta_{k+1},\beta_k) \overleftarrow{\partial_{\beta_k}}\right] G_B(\beta_1,\beta)\overleftarrow{\partial_{\beta}}\\
\end{array}
\end{equation}
where the $\textnormal{\footnotesize{DBL}}$ naturally follows from the first-passage decomposition, and the $\textnormal{\footnotesize{SBL}}$ naturally follows from the last-passage decomposition.
\end{CorollaryAbsorbed}

We only provide this corollary to emphasise that the difference between single and double boundary layers is arbitrary: what looks like a first passage or double boundary layer from the point of view of $x$, looks like a last passage or single boundary layer from the point of view of $y$. In other words, single/double boundary layers are only as different as first/last passage decompositions; i.e. not that different. In \cite{BalianBloch1970}, the \textit{multiple reflection expansion} is derived from the from the \textit{ansatz} of a double boundary layer and this approach has persisted in e.g. \cite{BalianBloch1974,BalianDuplantier1977,BalianDuplantier1978,HanssonJaffe1983a,HanssonJaffe1983b,Bordag1999,Bordag2001,Bordag2002,Bordag2005,Maghrebi2011}.  

From the \textit{pair} of FP and LP series, it is obvious that $G_A$ is symmetric. But if only \textit{one} series is derived by an ansatz, then it is \textit{not} obvious that the absorbed Green function $G_A$ is symmetric, or that it satisfies the boundary conditions, as noted in \cite{HanssonJaffe1983b}. To show that the series is indeed symmetric, \cite{HanssonJaffe1983b} suggest a symmetrisation procedure which involves singular terms like this one
\begin{equation*}
\frac{1}{4}\Bigg[\oint d\beta_2 \oint d\beta_1\Bigg]\,G_B(y,\beta_2)\,\overrightarrow{\partial_{\beta_2}}G_B(\beta_2,\beta_1)\overleftarrow{\partial_{\beta_1}}\,G_B(\beta_1,x)=\infty
\end{equation*}
The difference is subtle, but a quantity that does exist, is the following:
\begin{equation*}
\frac{1}{4}\oint d\beta_2\; G_B(y,\beta_2)\overleftrightarrow{\partial_{\beta_2}}\;\oint d\beta_1\; G_B(\beta_2,\beta_1)\overleftrightarrow{\partial_{\beta_2}}G_B(\beta_1,x)
\end{equation*}
where the differential operator with arrows both ways works on two sides. The `symmetrisation' mistake of \cite{HanssonJaffe1983b} is inherited by e.g. \cite{Bordag2001}. The term that we propose is considered in detail in section \ref{section4}.

Apart from the equivalence of single and double boundary layers, we further wish to emphasise that our result relies only on the applicability of Green's second identity. The original paper \cite{BalianBloch1970} was subtitled `Three dimensional problem with smooth boundary surface' and subsequent literature also assumes smooth boundaries. 

The reflected Green function exists in $d\geq3$, if the domain is infinite. We do not present the result explicitly, but in that case the reflected Green function can also be written as \textit{either} a single or double boundary layer --- by integrating Proposition \ref{Proposition4} over an infinite time. 

\subsection{Appendix: three lemmas}
\label{Lemmas}

\begin{absorbedlemma}\label{Lemma1} \textnormal{\textbf{Pushing differential operators through integrals.}}
For a test-function $f$ and a regular boundary coordinate $\beta$, we have 
\begin{equation*}
\begin{array}{r@{\hspace{1mm}}r@{\hspace{1mm}}c@{\hspace{1mm}}l@{\hspace{1mm}}l}
\displaystyle\left\{\frac{1}{2}\overrightarrow{\partial_\beta}\right\}&\displaystyle\left(\int_s^t d\tau \underset{\partial D}\oint d\gamma\; B(\beta,t|\gamma,\tau)f(\gamma,\tau)\right)&&=&\displaystyle-\frac{1}{2}f(\beta,t)\\
&&&&\displaystyle+\int_s^t d\tau \underset{\partial D}\oint d\gamma\;\left\{\frac{1}{2}\overrightarrow{\partial_\beta}\right\}B(\beta,t|\gamma,\tau)f(\gamma,\tau)\\[3ex]
&\displaystyle\left(\int_s^t d\tau \underset{\partial D}\oint d\gamma\; f(\gamma,\tau)B(\gamma,\tau|\beta,s)\right)&\displaystyle\left\{\frac{1}{2}\overleftarrow{\partial_\beta}\right\}&=&
\displaystyle-\frac{1}{2}f(\beta,s)\\
&&&&\displaystyle+\int_s^t d\tau \underset{\partial D}\oint d\gamma\; f(\gamma,\tau)B(\gamma,\tau|\beta,s)\left\{\frac{1}{2}\overleftarrow{\partial_\beta}\right\}\\
\end{array}
\end{equation*}
\end{absorbedlemma}

\begin{reflectedlemma}\label{Lemma2}\textnormal{\textbf{Pushing limits through integrals.}}
For a test-function $f$ and a regular boundary coordinate $\beta$, we have 
\begin{equation*}
\begin{array}{r@{\hspace{1mm}}c@{\hspace{1mm}}l}
\displaystyle\lim_{x \to \beta} \int_s^t d\tau \underset{\partial D}\oint d\gamma\; f(\gamma,\tau)\left\{\frac{1}{2}\overrightarrow{\partial_\gamma}\right\}B(\gamma,\tau|x,s)&=&\displaystyle\frac{1}{2}f(\beta,s)+\int_s^t d\tau \underset{\partial D}\oint d\gamma\; f(\gamma,\tau)\left\{\frac{1}{2}\overrightarrow{\partial_\gamma}\right\}B(\gamma,\tau|\beta,s)\\
\displaystyle\lim_{y \to \beta} \int_s^t d\tau \underset{\partial D}\oint d\gamma\; B(y,t|\gamma,\tau) \left\{\frac{1}{2}\overleftarrow{\partial_\gamma}\right\}f(\gamma,\tau)&=&\displaystyle\frac{1}{2}f(\beta,t)+\int_s^t d\tau \underset{\partial D}\oint d\gamma\; B(\beta,t|\gamma,\tau) \left\{\frac{1}{2}\overleftarrow{\partial_\gamma}\right\} f(\gamma,\tau)
\end{array}
\end{equation*}
\end{reflectedlemma}

\begin{arrowlemma}\label{Lemma3}\textnormal{\textbf{Changing directions of arrows.}}
By the fundamental theorem of calculus, the STCs and the fact that a Dirac $\delta$-function on a regular part of the boundary picks up half a contribution, we have that
\begin{equation*}
\int_s^t d\tau \left(\frac{\partial}{\partial \tau }\right)\underset{D}\int d\alpha\;  B(y,t|\alpha,\tau )B(\alpha,t|x,s)=
\left
\{
\begin{array}{llll}
 0 & \mbox{if} & x \in D, & y \in D; \\
 \frac{1}{2}B(y,t|x,s) & \mbox{if} & x \in \partial D^r, & y \in D; \\
 -\frac{1}{2}B(y,t|x,s) & \mbox{if} & x \in D, & y \in \partial D^r. 
\end{array} 
\right.
\end{equation*}
By the PDEs and Green's theorem, this in turn implies that
\begin{equation*}
\label{ChangingArrows}
\int_s^t d\tau \underset{\partial D}\oint d\beta\; B(y,t|\beta,\tau)\left\{\overleftarrow{\partial_\beta}-\overrightarrow{\partial_\beta}\right\}B(\beta,\tau|x,s) = 
\left
\{
\begin{array}{llll}
 0 & \mbox{if} & x \in D, & y \in D; \\
 B(y,t|x,s) & \mbox{if} & x \in \partial D^r, & y \in D; \\
 -B(y,t|x,s) & \mbox{if} & x \in D, & y \in \partial D^r. 
\end{array} \right.
\end{equation*}
\end{arrowlemma}

\section{Brownian motion and path integrals}
\label{section3}

\subsection{The Schr\"{o}dinger equation in a probabilistic setting}

In quantum mechanics the motion of a particle is determined by the Schr\"{o}dinger equation. We will transform the Schr\"{o}dinger equation into a probabilistic setting by going to imaginary time ($t\rightarrow-i\,t$). Larger mass $m$ of a particle (i.e. higher inertia) is analogous to lower variance $\sigma ^2$ of a Brownian motion, suggesting we set $m=\frac{1}{\sigma^2}$. With these changes and with $\hbar=1$, our version of the Schr\"{o}dinger equation reads as follows:
\begin{equation}
\label{SchrodingerEquation}
\begin{array}{rr@{\hspace{1mm}}c@{\hspace{1mm}}l}
\text{\scriptsize{forward PDE}}&\displaystyle\Big( \frac{\sigma^2}{2}\nabla_y^2-\frac{\partial}{\partial t}-\lambda\, V(y)\Big)\psi_V(y,t|x,s)&=&0\\[1.5ex]
\text{\scriptsize{backward PDE}}&\displaystyle\Big(\frac{\sigma^2}{2}\nabla_x^2+\frac{\partial}{\partial s}-\lambda\, V(x)\Big)\psi_V(y,t|x,s)&=&0\\[1.5ex]
\text{\scriptsize{forward BC}}&\displaystyle\underset{|y| \to  \infty}\lim\psi_V(y,t|x,s)&=&0\\[1.5ex]
\text{\scriptsize{backward BC}}&\displaystyle\underset{|x| \to  \infty}\lim\psi_V(y,t|x,s)&=&0\\[1.5ex]
\text{\scriptsize{forward STC}}&\displaystyle\lim_{s \nearrow t} \psi_V(y,t|x,s)&=&\delta(|y-x|)\\[1.5ex]
\text{\scriptsize{backward STC}}&\displaystyle\lim_{t \searrow s} \psi_V(y,t|x,s)&=&\delta(|y-x|)\\
\end{array}
\end{equation}
where PDE stands for partial differential equation, BC stands for boundary condition and STC stands for short-time condition. The coordinates $(y,t)$ and $(x,s)$ are referred to as the `forward' and `backward' space-time coordinates, respectively. We use the symbol $\psi$ since this is customary in quantum mechanics, but in our case $\psi$ is a probability density, where the dependence on the potential is indicated through the subscript. The  \textit{coupling constant} $\lambda$ measures the `strength' of the coupling with the potential $V$. 

The PDEs can be seen to hold through the following probabilistic interpretation. Suppose that we have a Brownian motion as before, except we add the possibility that some catastrophic event happens, during time $ds$, annihilating the particle and reducing to zero the probability of propagation to any location, at any later time. This event we call an \textit{interaction} with the potential. Suppose that an interaction happens with a probability that is a product of the strength of the potential at a certain location, and the time spent there. This means that during $ds$, and at location $x$, an interaction happens with probability $\lambda\,V(x)\,ds$. The transition density $\psi_V$ must be unbiased, and the `catastrophic event' interpretation implies that we must have 
\begin{equation*}
\psi_V(y,t|x,s)= (1-\lambda\,V(x)\,ds)\; \mathbb{E}\; \psi_V(y,t|x+dB,s+ds)+\lambda\,V(x)\,ds\times 0
\end{equation*} 
where with probability $(1-\lambda\,V(x)\,ds)$ the particle stays alive and where with probability $\lambda\,V(x)\,ds$ the particle gets annihilated by the potential. Using  It\^{o}'s lemma we obtain to first order in $ds$ that
\begin{equation*}
\Big(\frac{\sigma ^2}{2}\nabla _x^2+\frac{\partial }{\partial s}-\lambda\,V(x) \Big)\psi_V(y,t|x,s)=0
\end{equation*} 
and similarly for the forward PDE. If the Brownian particle is not annihilated but instead another Brownian particle is created, then with probability $\lambda\,V(x)\,ds$ the propagation density \textit{doubles}, i.e. 
\begin{equation*}
\begin{array}{r@{\hspace{1mm}}c@{\hspace{1mm}}l}
\psi_V(y,t|x,s)&=&(1-\lambda\,V(x)\,ds)\,\mathbb{E}\,\psi_V(y,t|x+dB,s+ds)\\
&&\quad+\lambda\,V(x)\,ds \times 2 \mathbb{E}\,\psi_V(y,t|x+dB,s+ds)\\[2ex]
&=&(1+\lambda\,V(x)\,ds)\,\mathbb{E}\, \psi_V(y,t|x+dB,s+ds)
\end{array}
\end{equation*}
which, to first order in $ds$, leads by It\^{o}'s lemma to 
\begin{equation*}
\Big(\frac{\sigma ^2}{2}\nabla _x^2+\frac{\partial }{\partial s}+\lambda\,V(x) \Big)\psi_V(y,t|x,s)=0
\end{equation*} 
This could have been obtained immediately by switching the \textit{sign} of $V$. Thus positive potentials annihilate particles, while negative potentials create particles.

The STCs are satisfied, finally, because the probability of an interaction is proportional to $ds$ and thus within a very short period of time, the Brownian particle stays 1) alive and 2) where it is.

\subsection{First- and last-interaction decompositions}

The original research on the Schr\"{o}dinger equation starts here. In this section we will think of $V$ as positive, i.e. annihilating particles. Then $\psi_V(y,t|x,s)$ represents the probability that the particle moves from $(x,s)$ to $(y,t)$ with zero interactions, i.e. without being annihilated by the potential. Using the STCs and the fundamental theorem of calculus, we can write down the following first-interaction (FI) and last-interaction (LI) decompositions:
\begin{equation}
\makebox[13cm][l]
{
$
\begin{array}{lr@{\hspace{1mm}}c@{\hspace{1mm}}l@{\hspace{1mm}}l}
\text{\tiny{FI}}&\psi_V(y,t|x,s)&=&\displaystyle B(y,t|x,s)+\int _s^td\tau\;  &\displaystyle\frac{\partial
}{\partial \tau }\underset{\mathbb{R}^d}{\int } d\alpha\;B(y,t|\alpha,\tau )\,\psi _V(\alpha ,\tau |x,s)\\
\text{\tiny{LI}}&\psi _V(y,t|x,s)&=&\displaystyle B(y,t|x,s)-\int _s^td\tau &\displaystyle\frac{\partial }{\partial\tau}\underset{\mathbb{R}^d}{\int } d\alpha\; \psi_V(y,t|\alpha,\tau)\,B(\alpha ,\tau |x,s)
\end{array}
$
}
\end{equation}
where $B$ equals the free Brownian propagator. The FI and LI decompositions hold by the virtue of the fundamental theorem of calculus and the STCs, i.e. first we have
\begin{equation}
\makebox[13cm][l]
{
$
\begin{array}{lr@{\hspace{1mm}}c@{\hspace{1mm}}l}
\text{\tiny{FI}}&\psi_V(y,t|x,s)&=&\displaystyle B(y,t|x,s)+\left(\underset{\tau \nearrow t}\lim-\underset{\tau \searrow s}\lim\right)\underset{\mathbb{R}^d}{\int } d\alpha\;B(y,t|\alpha,\tau )\,\psi _V(\alpha ,\tau |x,s)\\
\text{\tiny{LI}}&\psi _V(y,t|x,s)&=&\displaystyle B(y,t|x,s)-\left(\underset{\tau \nearrow t}\lim-\underset{\tau \searrow s}\lim\right)\underset{\mathbb{R}^d}{\int } d\alpha\; \psi_V(y,t|\alpha,\tau)\,B(\alpha ,\tau |x,s)
\end{array}
$
}
\end{equation}
and the STCs show that both decompositions hold. To explain the nomenclature, we introduce the first-interaction time $\tau ^{\text{\tiny{FI}}}$ and last-interaction time $\tau ^{\text{\tiny{LI}}}$. They are defined as the first and last times that an interaction happens in the interval $[s,t]$. We use the conventions $\inf\{\varnothing\}=\infty$ and $\sup\{\varnothing\}=-\infty$ such that e.g. $\tau ^{\text{\tiny{FI}}}$ is infinite if no interaction happens in the interval $[s,t]$. With these definitions we have
\begin{equation}
\makebox[13cm][l]
{
$
\begin{array}{lr@{\hspace{1mm}}c@{\hspace{1mm}}l}
\text{\tiny{FI}}&\mathbb{P}\left(B_t \in dy\text{; }\tau ^{\text{\tiny{FI}}}\geq\tau\big|B_s=x\right)&=&\displaystyle\underset{\mathbb{R}^d}\int d\alpha\;  B(y,t|\alpha,\tau)\,\psi_V(\alpha,\tau|x,s)\\[2ex]
\text{\tiny{LI}}&\mathbb{P}\left(B_t \in dy\text{; }\tau ^{\text{\tiny{LI}}}\leq\tau\big|B_s=x\right)&=&\displaystyle\underset{\mathbb{R}^d}{\int} d\alpha\;  \psi_V(y,t|\alpha,\tau)\,B(\alpha,\tau|x,s)
\end{array}
$
}
\end{equation}
Differentiating, we find
\begin{equation}
\makebox[13cm][l]
{
$
\begin{array}{lr@{\hspace{1mm}}c@{\hspace{1mm}}r@{\hspace{1mm}}l}
\text{\tiny{FI}}&\mathbb{P}\left(B_t \in dy\text{; }\tau ^{\text{\tiny{FI}}}\in d\tau\big|B_s=x\right)&=&\displaystyle-\frac{\partial
}{\partial \tau }&\displaystyle\underset{\mathbb{R}^d}{\int} d\alpha\;  B(y,t|\alpha,\tau)\,\psi_V(\alpha,\tau|,s)\\[2ex]
\text{\tiny{LI}}&\mathbb{P}\left(B_t \in dy\text{; }\tau ^{\text{\tiny{LI}}}\in d\tau\big|B_s=x\right)&=&\displaystyle\frac{\partial
}{\partial \tau }&\displaystyle \underset{\mathbb{R}^d}{\int} d\alpha\;  \psi_V(y,t|\alpha,\tau)\,B(\alpha,\tau|,s)
\end{array}
$
}
\end{equation}
Then we subtract from the free density an integral (over $\tau$) over all paths with a first or last interaction at time $\tau$, and obtain
\begin{equation}
\makebox[13cm][l]
{
$
\begin{array}{lr@{\hspace{1mm}}c@{\hspace{1mm}}l@{\hspace{1mm}}c@{\hspace{1mm}}l}
\text{\tiny{FI}}&\psi_V(y,t|x,s)&=&\displaystyle B(y,t|x,s)+\int_s^t d\tau &\displaystyle\frac{\partial
}{\partial \tau }&\displaystyle\underset{\mathbb{R}^d}{\int } d\alpha\;B(y,t|\alpha,\tau )\,\psi _V(\alpha ,\tau |x,s) \\[2ex]
\text{\tiny{LI}}&\psi_V(y,t|x,s)&=&\displaystyle B(y,t|x,s)-\int_s^t d\tau &\displaystyle\frac{\partial
}{\partial \tau }&\displaystyle\underset{\mathbb{R}^d}{\int } d\alpha\;\psi_V(y,t|\alpha,\tau )\,B(\alpha ,\tau |x,s) 

\end{array}
$
}
\end{equation}
And thus we have re-derived the set of identities through a probabilistic intuition. While the interpretation of first and last interactions presents itself naturally for a positive (i.e. killing) potential, it is obvious that both identities hold for any reasonably behaved potential, since they only rely on the STCs. Using the PDEs of \eqref{SchrodingerEquation}, we get
\begin{equation}
\makebox[13cm][l]
{
$
\begin{array}{lr@{\hspace{1mm}}c@{\hspace{1mm}}l}
\text{\tiny{FI}}&\psi_V(y,t|x,s)&=&\displaystyle B(y,t|x,s)\\[2ex]
&&&\displaystyle-\int_s^t d\tau \underset{\mathbb{R}^d}{\int } d\alpha\;B(y,t|\alpha,\tau)\left(\frac{\sigma^2}{2}\overleftarrow{\nabla}^2_\alpha-\frac{\sigma^2}{2}\overrightarrow{\nabla}^2_\alpha+\lambda\,V(\alpha)\right)\psi _V(\alpha ,\tau |x,s) \\[2ex]
\text{\tiny{LI}}&\psi_V(y,t|x,s)&=&\displaystyle B(y,t|x,s)\\[2ex]
&&&\displaystyle+\int_s^t d\tau \underset{\mathbb{R}^d}{\int } d\alpha\,\psi_V(y,t|\alpha,\tau )\left(\frac{\sigma^2}{2}\overleftarrow{\nabla}^2_\alpha-\frac{\sigma^2}{2}\overrightarrow{\nabla}^2_\alpha-\lambda\,V(\alpha)\right)B(\alpha ,\tau |x,s) 

\end{array}
$
}
\end{equation}
where the direction of the arrows indicates the direction of differentiation. Using Green's identity for the differentiated terms, we can transform the integral over the `interior' of $\mathbb{R}^d$ to one over the `boundary' of $\mathbb{R}^d$. The BCs demand that the boundary terms disappear, and thus we obtain:

\begin{Potential}\label{Proposition5} \textnormal{\textbf{FI and LI decomposition of $\bm{\psi_V}$.}} The following formulations of a Brownian motion, in the presence of a well-behaved potential $V$, are equivalent:
\begin{equation}
\label{FirtAndLastInteractionDecompositions}
\left.
\begin{array}{r}
\displaystyle\Big(\frac{\sigma^2}{2}\nabla_y^2-\frac{\partial}{\partial t} -\lambda\, V(y)\Big)\psi_V(y,t|x,s)=0\\[1.5ex]
\displaystyle\Big(\frac{\sigma^2}{2}\nabla_x^2+\frac{\partial}{\partial s} -\lambda\, V(x)\Big)\psi_V(y,t|x,s)=0\\[1.5ex]
\displaystyle\underset{|y| \to \infty}\lim\psi_V(y,t|x,s)=0\\[1.5ex]
\displaystyle\underset{|x| \to \infty}\lim\psi_V(y,t|x,s)=0\\[1.5ex]
\displaystyle\lim_{s \nearrow t} \psi_V(y,t|x,s)=\delta(|y-x|)\\[1.5ex]
\displaystyle\lim_{t \searrow s} \psi_V(y,t|x,s)=\delta(|y-x|)\\
\end{array}\right\}=\left\{
\begin{array}{ll}
\textnormal{\tiny{FI}}&\psi_V(y,t|x,s)=\displaystyle B(y,t|x,s)\\
&\displaystyle-\int_s^t d\tau \underset{\mathbb{R}^d}{\int } d\alpha\;B(y,t|\alpha,\tau)\,\lambda\,V(\alpha)\,\psi _V(\alpha ,\tau |x,s)\\[4ex]
\textnormal{\tiny{LI}}&\psi_V(y,t|x,s)=\displaystyle B(y,t|x,s)\\
&\displaystyle-\int_s^t d\tau \underset{\mathbb{R}^d}{\int } d\alpha\;\psi_V(y,t|\alpha,\tau)\,\lambda\,V(\alpha)\,B(\alpha ,\tau |x,s)
\end{array}
\right.
\end{equation}
\end{Potential}
To physicists, the integral equations on the right-hand side are sometimes known as the Lippmann-Schwinger or Dyson equations, but our derivation and interpretation are different. By a repeated substitution of the equations into themselves, the solution may be written as
\begin{equation}
\begin{array}{r@{\hspace{1mm}}c@{\hspace{1mm}}l}
\psi_V(y,t|x,s)&=&\displaystyle B(y,t|x,s)+\sum_{i=1}^\infty\; (-\lambda)^i\, (K*)^i\, B(y,t|x,s)\\
\psi_V(y,t|x,s)&=&\displaystyle B(y,t|x,s)+\sum_{i=1}^\infty\; (-\lambda)^i\, B(y,t|x,s)\, (*K)^i
\end{array}
\end{equation}
where the operator $K$ is defined as follows
\begin{equation}
\label{IntegralOperatorK}
\begin{array}{r@{\hspace{1mm}}c@{\hspace{1mm}}l}
K*f(y,t|x,s)&:=&\displaystyle\int _s^td\tau\underset{\mathbb{R}^d}{\int} d\alpha\; B(y,t|\alpha
,\tau)\,V(\alpha)\,f(\alpha,\tau|x,s)\\
f(y,t|x,s)*K&:=&\displaystyle\int_s^td\tau\underset{\mathbb{R}^d}{\int}d\alpha\; f(y,t|\alpha,\tau
)\,V(\alpha)\,B(\alpha,\tau|x,s)
\end{array}
\end{equation}

\subsection{Decomposition for singular potentials}
\label{subsection3.3}

Let us introduce a potential $V_\epsilon(x)$ that is non-singular for all $\epsilon>0$. We are interested in the limit where $\epsilon$ goes down to zero. We define $\psi_{\epsilon=0}$ as follows:
\begin{equation}
\begin{array}{r@{\hspace{1mm}}c@{\hspace{1mm}}l}
\psi_\epsilon(y,t|x,s)&:=&\psi_{V_\epsilon}(y,t|x,s)\\
\psi_{\epsilon=0}(y,t|x,s)&:=&\underset{\epsilon \searrow 0}\lim\,\psi_{\epsilon}(y,t|x,s)
\end{array}
\end{equation} 
Not for all singular potentials $V_\epsilon$ does the limit $\psi_{\epsilon=0}(y,t|x,s)$ exist. But there are singular potentials $V_\epsilon$ for which the limit $\psi_{\epsilon=0}(y,t|x,s)$ does exist, at least for $x$ and $y$ away from the singularity of $V_\epsilon$. We will assume that the limit $\psi_{\epsilon=0}$ indeed exists and proceed from there. For all $\epsilon>0$, the FI and LI decompositions of Proposition \ref{Proposition5} should apply, i.e.  
\begin{equation}
\makebox[13cm][l]
{
$
\begin{array}{lr@{\hspace{1mm}}c@{\hspace{1mm}}l}
\text{\tiny{FI}}&\psi_{\epsilon}(y,t|x,s)&=&\displaystyle B(y,t|x,s)-\int_s^t d\tau \underset{\mathbb{R}^d}{\int } d\alpha\;B(y,t|\alpha,\tau)\,\lambda\,V_\epsilon(\alpha)\,\psi _\epsilon(\alpha ,\tau |x,s)\\[2ex]
\text{\tiny{LI}}&\psi_\epsilon(y,t|x,s)&=&\displaystyle B(y,t|x,s)-\int_s^t d\tau \underset{\mathbb{R}^d}{\int } d\alpha\;\psi_\epsilon(y,t|\alpha,\tau)\,\lambda\,V_\epsilon(\alpha)\,B(\alpha ,\tau |x,s)

\end{array}
$
}
\end{equation}
Applying to both sides the limit where $\epsilon$ goes to zero, we obtain:
\begin{equation}
\makebox[13cm][l]
{
$
\begin{array}{lr@{\hspace{1mm}}c@{\hspace{1mm}}l}
\text{\tiny{FI}}&\displaystyle\psi_{\epsilon=0}(y,t|x,s)&=&\displaystyle B(y,t|x,s)-\underset{\epsilon \searrow 0}\lim\int_s^t d\tau \underset{\mathbb{R}^d}{\int } d\alpha\;B(y,t|\alpha,\tau)\,\lambda\,V_\epsilon(\alpha)\,\psi _\epsilon(\alpha ,\tau |x,s)\\[2ex]
\text{\tiny{LI}}&\displaystyle\psi_{\epsilon=0}(y,t|x,s)&=&\displaystyle B(y,t|x,s)-\underset{\epsilon \searrow 0}\lim\int_s^t d\tau \underset{\mathbb{R}^d}{\int } d\alpha\;\psi_\epsilon(y,t|\alpha,\tau)\,\lambda\,V_\epsilon(\alpha)\,B(\alpha ,\tau |x,s)

\end{array}
$
}
\end{equation}
We cannot push the limit through the integrals on the right-hand side, because the potential $V_\epsilon$ is singular in the limit where $\epsilon$ goes to zero. But the quantity $\psi_\epsilon$, which also appears under the integral on the right-hand side, should remain well-behaved --- by assumption. Using that the limit of a product is equal to the product of limits, we can push the limit regarding $\psi_\epsilon$ through the integral sign, and obtain the following Proposition:

\begin{SingularPotential}\label{Proposition6} \textnormal{\textbf{Decomposition for singular potentials.}} If it exists, the propagator $\psi_{\epsilon=0}$ satisfies the following decompositions:
\begin{equation}
\label{singularpotentials}
\makebox[13cm][l]
{
$
\begin{array}{lr@{\hspace{1mm}}c@{\hspace{1mm}}l}
\textnormal{\tiny{FI}}&\displaystyle\psi_{\epsilon=0}(y,t|x,s)&=&\displaystyle B(y,t|x,s)-\underset{\zeta \searrow 0}\lim\int_s^t d\tau \underset{\mathbb{R}^d}{\int } d\alpha\;B(y,t|\alpha,\tau)\,\lambda\,V_{\zeta}(\alpha)\,\psi_{\epsilon=0}(\alpha ,\tau |x,s),\\[2ex]
\textnormal{\tiny{LI}}&\displaystyle\psi_{\epsilon=0}(y,t|x,s)&=&\displaystyle B(y,t|x,s)-\underset{\zeta \searrow 0}\lim\int_s^t d\tau \underset{\mathbb{R}^d}{\int } d\alpha\;\psi_{\epsilon=0}(y,t|\alpha,\tau)\,\lambda\,V_{\zeta}(\alpha)\,B(\alpha ,\tau |x,s).

\end{array}
$
}
\end{equation}
\end{SingularPotential}
\noindent These equations may be written as
\begin{equation}
\makebox[10cm][l]
{
$
\begin{array}{lr@{\hspace{1mm}}c@{\hspace{1mm}}l}
\textnormal{\tiny{FI}}&\psi_{\epsilon=0}(y,t|x,s)&=&\displaystyle B(y,t|x,s)-\lambda\, L*\psi_{\epsilon=0}(y,t|x,s)\\[2ex]
\textnormal{\tiny{LI}}&\displaystyle\psi_{\epsilon=0}(y,t|x,s)&=&\displaystyle B(y,t|x,s)-\lambda\, \psi_{\epsilon=0}(y,t|x,s)*L
\end{array}
$
}
\end{equation}
provided that we introduce the operator $L$ as follows
\begin{equation}
\label{Loperator}
\begin{array}{r@{\hspace{1mm}}c@{\hspace{1mm}}l}
L*g(y,t|x,s)&:=&\displaystyle\underset{\zeta \searrow 0}\lim\int _s^td\tau\underset{\mathbb{R}^d}{\int} d\alpha\; B(y,t|\alpha
,\tau)\,V_\zeta(\alpha)\,g(\alpha,\tau|x,s)\\
g(y,t|x,s)*L&:=&\underset{\zeta \searrow 0}\lim\displaystyle\int_s^td\tau\underset{\mathbb{R}^d}{\int}d\alpha\; g(y,t|\alpha,\tau
)\,V_\zeta(\alpha)\,B(\alpha,\tau|x,s)
\end{array}
\end{equation}
And by a repeated substitution, the following series solution is obtained:
\begin{equation}
\label{seriesforsingularpotential}
\begin{array}{r@{\hspace{1mm}}c@{\hspace{1mm}}l}
\psi_{\epsilon=0}(y,t|x,s)&=&\displaystyle B(y,t|x,s)+\sum_{i=1}^\infty\; (-\lambda)^i\, (L*)^i\, B(y,t|x,s)\\
\psi_{\epsilon=0}(y,t|x,s)&=&\displaystyle B(y,t|x,s)+\sum_{i=1}^\infty\; (-\lambda)^i\, B(y,t|x,s)\, (*L)^i
\end{array}
\end{equation}
As an example of a well-known singular potential, we consider the propagator for the one-dimensional Dirac $\delta$-potential. It can be found in e.g. \cite[p.~381]{Schulman1981techniques}, and is as follows
\begin{equation}
\label{deltasolution}
\psi(y,t|x,s)=B(y,t|x,s)-\lambda \int_0^\infty d\alpha\; e^{-\lambda \alpha}\,B(|y|+|x|+\alpha,t|0,s)
\end{equation}
It can be checked by direct calculation that it satisfies the integral equations \eqref{singularpotentials}, when the potential $V$ is taken to be a Gaussian, for example, that approaches the Dirac $\delta$-function. Consequently, the series solution \eqref{seriesforsingularpotential} converges in an alternating fashion to the exact solution \eqref{deltasolution}. For the Dirac $\delta$-function, both the solution and all correction terms are continuous at zero. Therefore it makes no difference if the Dirac $\delta$-function is taken to be symmetric or not. 
 
\subsection{The Feynman-Kac formula}

The Feynman-Kac formula, which appeared in \cite{Kac1949}, suggests itself through the interpretation of $V$ as as a rate of killing. It is useful because it allows us to write the series solution of Proposition \ref{Proposition5} in a very compact manner. We will only derive it heuristically, for a more formal treatment see e.g. \cite{MorterPeres2010}.

We slice the time from $s$ to $t$ up such that there are $N$ intermediate locations. The length of each time interval is $\epsilon=(t-s)/(N+1)$. Using the subscript $i$ to indicate time, we define $\tau _i= s+i\,\epsilon$. Let $i$ run from 0 to $N+1$ such that $\tau _0=s$ and $\tau_{N+1}=t$. Thus there are $N$ intermediate times. The path from $(x,s)$ to $(y,t)$ is defined by $N$ intermediate locations $\{B_{\tau _1},\ldots,B_{\tau _N}\}$. The probability of survival of this path is a product of $N+1$ probabilities: one for each intermediate location, and one for the end-point (it is assumed that the particle is not annihilated at the starting point). Therefore the probability of survival equals
\begin{equation}
\prod _{i=1}^{N+1}\left(1-\lambda\, V(B_{\tau _i})\,\epsilon\right)\approx\prod_{i=1}^{N+1} e^{-\lambda\, V(B_{\tau _i})\,\epsilon}=e^{-\lambda  \sum_{i=1}^{N+1} V(B_{\tau _i})\,\epsilon}\to e^{-\lambda\int_s^t V(B_{\tau}) d\tau}
\end{equation} 
and where the last relationship holds in the limit for large $N$. If the above is the probability that a \textit{given} path should survive (with $N$ known intermediate locations), then the probability that \textit{any} path should survive is obtained by taking an expectation over all possible intermediate locations, i.e. over all paths. If we want the path to end up at $y$ then we need to take an expectation over all paths while enforcing the last position to be $y$. We can achieve this by plugging in a $\delta$-function at $y$. With this heuristic, we are lead to propose the \textit{Feynman-Kac formula} as follows:
\begin{equation}
\psi_V(y,t|x,s)=\mathbb{E}_x\,\Bigg[\delta(B_t-y)\,\exp\Big[-\lambda\,\int_s^t V(B_\tau)\,d\tau \Big]\Bigg]
\end{equation} 
We note that a positive potential, which kills paths, leads to a propagator $\psi _V$ which is smaller than the free propagator $B$. A negative potential, which creates paths, causes $\psi _V$ to be larger than the free density $B$.

\section{The Laplacian of the indicator}
\label{section4}

The first/last interaction decompositions of section \ref{section3} are very similar to the first/last passage decompositions of section \ref{section2}. In this section, we show that those sections are not merely similar, but equivalent, if we postulate a particular singular potential $V$.

\subsection{Differentiating the indicator}
\label{differentiating}

First, recall the fundamental theorem of calculus:
\begin{equation}
\int_a^b \frac{\partial f(x)}{\partial x}\,dx=\underset{x \nearrow b}\lim f(x)-\underset{x \searrow a}\lim f(x)
\end{equation}
for a trial function $f$ that is defined in the interval $(a,b)$, i.e. it does not need to be defined outside the interval. Now take $a<0$ and $b>0$. Then we have that
\begin{equation}
\begin{array}{r@{\hspace{1mm}}c@{\hspace{1mm}}l}
\displaystyle\int_a^b f(x) \frac{\partial \mathbbm{1}_{x>0}}{\partial x}  \,dx &=& \displaystyle\int_a^b \frac{\partial}{\partial x} \big(f(x) \mathbbm{1}_{x>0}\big) \,dx - \int_a^b \frac{\partial f(x)}{\partial x}  \mathbbm{1}_{x>0} \,dx\\[2ex]
&=&\displaystyle\underset{x \nearrow b}\lim f(x) - \int_0^b \frac{\partial f(x)}{\partial x}  \,dx\\[2ex]
&=&\displaystyle\underset{x \searrow 0}\lim f(x)
\end{array}
\end{equation}
Even though derivatives of the indicator function do not exist at zero, following the usual rules of partial integration produces the `correct' answer. Notice how it gives the value of $f$ just to the right of zero. If we take a bump function $I_\epsilon(x)$ that approaches the indicator $\mathbbm{1}_{x>0}$ from below, then 
\begin{equation*}
\begin{array}{r@{\hspace{1mm}}c@{\hspace{1mm}}l}
\displaystyle\underset{\epsilon \searrow 0}\lim\int_{a}^{b}\, f(x) \frac{\partial I_\epsilon(x)}{\partial x}\;dx&=&\displaystyle\underset{x \searrow 0}\lim f(x)\\
\end{array}
\end{equation*}
This result agrees with what we obtained above, by a naive application of an integration by parts, but only because we have chosen our bump function to approach the indicator from below. We now turn to double derivatives. In one dimension, for $a<b$, we may write
\begin{equation}
\begin{array}{r@{\hspace{1mm}}c@{\hspace{1mm}}l}
\displaystyle\int_{-\infty}^{+\infty}  \frac{\partial^2\mathbbm{1}_{a<x<b}}{\partial x^2}\,f(x)\;dx&=&\displaystyle\int_{-\infty}^{+\infty}  \mathbbm{1}_{a<x<b} \frac{\partial^2 f(x)}{\partial x^2}\;dx\\
&=&\displaystyle\int_{a}^{b}\frac{\partial^2 f(x)}{\partial x^2}\;dx\\
&=&\displaystyle\Big(\underset{ x \nearrow b}\lim -\underset{ x \searrow a}\lim\Big) \frac{\partial f(x)}{\partial x}
\end{array}
\end{equation}
where the first equality follows from the fact that two integrations by parts yield no boundary terms, because $\mathbbm{1}_{a<x<b}$ and $\partial_x \mathbbm{1}_{a<x<b}$ both vanish at infinity. The third line can be seen as a `sum' of `outward normal derivatives' --- where the `sum' is over two boundary locations. This sum becomes an integral in higher dimensions, which we show as follows:
\begin{equation}
\begin{array}{r@{\hspace{1mm}}c@{\hspace{1mm}}l}
\displaystyle \int _{\mathbb{R}^d}\nabla_x^2\mathbbm{1}_{x\in D}\,f(x)\;dx&=&\displaystyle \int _{\mathbb{R}^d}\mathbbm{1}_{x\in D}\,\nabla_x^2 f(x)\;dx\\[2ex]
&=&\displaystyle \int _{D}\,\nabla_x^2 f(x)\;dx\\[2ex]
&=&\displaystyle \oint _{\partial D}\,\underset{\alpha \to \beta}\lim n_\beta \cdot \nabla_\alpha f(\alpha)\;d\beta
\end{array}
\end{equation}
where the first equality follows from Green's second identity and the fact that $\mathbbm{1}_{x\in D}$ as well as $\nabla _x\mathbbm{1}_{x\in D}$ are zero when evaluated at the `boundary' of $\mathbb{R}^d$. Just like in the one-dimensional case, we get a sum (or integral) over the normal derivative at all boundary locations. For a finite domain (or when $f$ vanishes at infinity), we obtain by the divergence theorem 
\begin{equation*}
\begin{array}{r@{\hspace{1mm}}c@{\hspace{1mm}}l}
\displaystyle \int _{\mathbb{R}^d}\nabla_x^2\big[\mathbbm{1}_{x\in D}\,f(x)\big]\;dx&=&0,\\
\displaystyle \int _{\mathbb{R}^d}\,\nabla_x^2\mathbbm{1}_{x\in D}\,f(x)\;dx+ \displaystyle \int _{\mathbb{R}^d}\mathbbm{1}_{x\in D}\,\nabla_x^2 f(x)\;dx
&=&-2 \displaystyle \int _{\mathbb{R}^d}\nabla_x\mathbbm{1}_{x\in D}\cdot \nabla_x f(x)\;dx,\\
\displaystyle \oint_{\partial D}\,\underset{\alpha \to \beta}\lim n_\beta \cdot \nabla_\alpha f(\alpha)\;d\beta &=& - \displaystyle \int _{\mathbb{R}^d}\nabla_x\mathbbm{1}_{x\in D}\cdot \nabla_x f(x)\;dx.
\end{array}
\end{equation*}
The third equality follows from our previous analysis. We may choose $f$ such that $\nabla_x f(x)$ behaves like $n_x\,f(x)$ near the boundary. In this case we obtain  
\begin{equation}
\label{multidimdelta}
\begin{array}{r@{\hspace{1mm}}c@{\hspace{1mm}}l}
\displaystyle \oint _{\partial D}\,\underset{\alpha \to \beta}\lim f(\alpha)\;d\beta = - \displaystyle \int _{\mathbb{R}^d}\,f(x)\, n_x\cdot \nabla_x\mathbbm{1}_{x\in D}\;dx,
\end{array}
\end{equation}
where $\alpha$ moves to the boundary point $\beta$ from the inside of $D$. This shows that $-n_x\cdot \nabla_x\mathbbm{1}_{x\in D}$ is the proper generalisation of the inward normal derivative $\partial_x  \mathbbm{1}_{{x}>0}$ and the Dirac $\delta$-function in one dimension. 

\subsection{ABM and EBM by potentials}
\label{ABMpotential}

In this subsection we will show how to write the integral equations of ABM and EBM as caused by a potential. Recall that, by virtue of the STCs alone, for ABM we have the following pair of identities:
\begin{equation}
\makebox[13cm][l]
{
$
\begin{array}{lr@{\hspace{1mm}}c@{\hspace{1mm}}l@{\hspace{1mm}}l@{\hspace{1mm}}r@{\hspace{1mm}}ll}
\text{\tiny{FP}}&A(y,t|x,s)&=&B(y,t|x,s)&+&\displaystyle\int _s^td\tau &\displaystyle\frac{\partial}{\partial{\tau}}&\displaystyle\underset{D}{\int } d\alpha\;B(y,t|\alpha,\tau)A(\alpha,\tau|x,s)\\
\text{\tiny{LP}}&A(y,t|x,s)&=&B(y,t|x,s)&-&\displaystyle\int _s^td\tau  &\displaystyle\frac{\partial}{\partial{\tau}}&\displaystyle\underset{D}{\int} d\alpha\;A(y,t|\alpha,\tau)B(\alpha,\tau|x,s)
\end{array}
$
}
\end{equation}
As usual, by using the PDEs we obtain:
\begin{equation}
\makebox[13cm][l]
{
$
\begin{array}{lr@{\hspace{1mm}}c@{\hspace{1mm}}l}
\text{\tiny{FP}}&A(y,t|x,s)&=&B(y,t|x,s)-\displaystyle\frac{\sigma ^2}{2}\int _s^td\tau \underset{D}{\int} d\alpha\;B(y,t|\alpha,\tau)\left \{\overleftarrow\nabla_\alpha^2-\overrightarrow\nabla_\alpha^2\right\}A(\alpha,\tau|x,s)\\
\text{\tiny{LP}}&A(y,t|x,s)&=&B(y,t|x,s)+\displaystyle\frac{\sigma ^2}{2}\int _s^td\tau  \underset{D}{\int} d\alpha\;A(y,t|\alpha,\tau)\left\{\overleftarrow\nabla_\alpha^2-\overrightarrow\nabla_\alpha^2\right\}B(\alpha,\tau|x,s)
\end{array}
$
}
\end{equation}
Proceeding as before, we use Green's second identity to obtain
\begin{equation}
\makebox[13cm][l]
{
$
\begin{array}{lr@{\hspace{1mm}}c@{\hspace{1mm}}l}
\text{\tiny{FP}}&A(y,t|x,s)&=&B(y,t|x,s)+\displaystyle\frac{1}{2}\int _s^td\tau \underset{\partial D}{\oint} d\beta\,\,B(y,t|\beta,\tau)\left\{\overleftarrow{\partial_\beta}-\overrightarrow{\partial_\beta}\right\}A(\beta,\tau|x,s)\\
\text{\tiny{LP}}&A(y,t|x,s)&=&B(y,t|x,s)-\displaystyle\frac{1}{2}\int _s^td\tau  \underset{\partial D}{\oint} d\beta\,\,A(y,t|\beta,\tau)\left\{\overleftarrow{\partial_\beta}-\overrightarrow{\partial_\beta}\right\}B(\beta,\tau|x,s)\\
\end{array}
$
}
\end{equation}
where $\partial_\beta$ is again the scaled inward normal derivative. The BCs of require that $A$ is zero on the boundary, or at least on all regular parts. Instead of discarding the boundary terms that vanish by the BCs, we may \textit{change their sign} to obtain:
\begin{equation}
\makebox[13cm][l]
{
$
\begin{array}{lr@{\hspace{1mm}}c@{\hspace{1mm}}l}
\text{\tiny{FP}}&A(y,t|x,s)&=&B(y,t|x,s)-\displaystyle\frac{1}{2}\int _s^td\tau \underset{\partial D}{\oint} d\beta\,\,B(y,t|\beta,\tau)\left\{\overleftarrow{\partial_\beta}+\overrightarrow{\partial_\beta}\right\}A(\beta,\tau|x,s)\\
\text{\tiny{LP}}&A(y,t|x,s)&=&B(y,t|x,s)-\displaystyle\frac{1}{2}\int _s^td\tau  \underset{\partial D}{\oint} d\beta\,\,A(y,t|\beta,\tau)\left\{\overleftarrow{\partial_\beta}+\overrightarrow{\partial_\beta}\right\}B(\beta,\tau|x,s)\\
\end{array}
$
}
\end{equation}
By the divergence theorem this equals
\begin{equation}
\makebox[13cm][l]
{
$
\begin{array}{lr@{\hspace{1mm}}c@{\hspace{1mm}}l}
\text{\tiny{FP}}&A(y,t|x,s)&=&B(y,t|x,s)+\displaystyle\frac{\sigma^2}{2}\int _s^td\tau \underset{D}{\int} d\alpha\,\,\nabla_\alpha^2 \bigg[B(y,t|\alpha,\tau)A(\alpha,\tau|x,s)\bigg]\\
\text{\tiny{LP}}&A(y,t|x,s)&=&B(y,t|x,s)+\displaystyle\frac{\sigma^2}{2}\int _s^td\tau  \underset{D}{\int} d\alpha\,\,\nabla_\alpha^2 \bigg[ A(y,t|\alpha,\tau)B(\alpha,\tau|x,s)\bigg]\\
\end{array}
$
}
\end{equation}
Nothing stops us from extending the integration over all of $\mathbb{R}^d$ as long as we also insert an indicator function $\mathbbm{1}_{\alpha \in D}$ into the integrand, i.e.
\begin{equation}
\label{integrand}
\makebox[13cm][l]
{
$
\begin{array}{lr@{\hspace{1mm}}c@{\hspace{1mm}}l}
\text{\tiny{FP}}&A(y,t|x,s)&=&B(y,t|x,s)-\displaystyle \int _s^td\tau \underset{\mathbb{R}^d}{\int} d\alpha\;\mathbbm{1}_{\alpha \in D} \left(-\frac{\sigma^2}{2}\nabla_\alpha^2\right) \bigg[B(y,t|\alpha,\tau)A(\alpha,\tau|x,s)\bigg]\\
\text{\tiny{LP}}&A(y,t|x,s)&=&B(y,t|x,s)-\displaystyle \int _s^td\tau  \underset{\mathbb{R}^d}{\int} d\alpha\;\mathbbm{1}_{\alpha \in D} \left(-\frac{\sigma^2}{2}\nabla_\alpha^2\right) \bigg[ A(y,t|\alpha,\tau)B(\alpha,\tau|x,s)\bigg]\\
\end{array}
$
}
\end{equation}
where no definition for $A$ is required outside of $D$ (and any definition is allowed), since the indicator is zero there. Both the indicator and its divergence vanish at the `boundary' of $\mathbb{R}^d$. Using Green's theorem where the boundary terms disappear, the Laplacian now operates on the indicator function as follows:
\begin{equation}
\makebox[13cm][l]
{
$
\begin{array}{lr@{\hspace{1mm}}c@{\hspace{1mm}}l}
\text{\tiny{FP}}&A(y,t|x,s)&=&B(y,t|x,s)-\displaystyle \int _s^td\tau \underset{\mathbb{R}^d}{\int} d\alpha\; B(y,t|\alpha,\tau)\bigg[-\frac{\sigma^2}{2}\nabla_\alpha^2 \mathbbm{1}_{\alpha \in D}\bigg]A(\alpha,\tau|x,s)\\
\text{\tiny{LP}}&A(y,t|x,s)&=&B(y,t|x,s)-\displaystyle \int _s^td\tau  \underset{\mathbb{R}^d}{\int} d\alpha\; A(y,t|\alpha,\tau)\bigg[-\frac{\sigma^2}{2}\nabla_\alpha^2\mathbbm{1}_{\alpha \in D}\bigg]B(\alpha,\tau|x,s)\\
\end{array}
$
}
\end{equation}
We have previously shown that a Brownian particle that is allowed in all of $\mathbb{R}^d$ but acted upon by a potential $V$ satisfies the FI and LI decompositions in Proposition \ref{Proposition5}. Comparing, we can associate the absorbing potential in the following way:
\begin{equation*}
V(\alpha):=- \frac{\sigma^2}{2}\,\nabla_{\alpha}^2 \mathbbm{1}_{\alpha\in D} 
\end{equation*}
The above may seem purely formal, but it does suggest the following limiting procedure:
\begin{equation}
\makebox[13cm][l]
{
$
\begin{array}{lr@{\hspace{1mm}}c@{\hspace{1mm}}l}
\text{\tiny{FP}}&A(y,t|x,s)&=&B(y,t|x,s)-\displaystyle \underset{\epsilon \searrow 0}\lim \int _s^td\tau \underset{\mathbb{R}^d}{\int} d\alpha\; B(y,t|\alpha,\tau)\bigg[-\frac{\sigma^2}{2} I''_{\epsilon}(\alpha)\bigg]A(\alpha,\tau|x,s)\\
\text{\tiny{LP}}&A(y,t|x,s)&=&B(y,t|x,s)-\displaystyle \underset{\epsilon \searrow 0}\lim \int _s^td\tau  \underset{\mathbb{R}^d}{\int} d\alpha\; A(y,t|\alpha,\tau)\bigg[-\frac{\sigma^2}{2}I''_{\epsilon}(\alpha)\bigg]B(\alpha,\tau|x,s)\\
\end{array}
$
}
\end{equation}
where we have used the short-hand $I''_{\epsilon}=\nabla^2 I_\epsilon$. Now we do have a prescription that is well-defined, and we may compare with the FI and LI decompositions in Proposition \ref{Proposition6}, which were specifically formulated for singular potentials, to check that the potential $V$ may be identified as follows:
\begin{equation*}
V(\alpha):=-\frac{\sigma^2}{2}\underset{\epsilon \searrow 0}\lim\,\nabla_\alpha^2 I_\epsilon(\alpha)
\end{equation*}
For the elastic density $E$ we can derive a similar result. We start with the following identities, which hold by the virtue of the STCs:
\begin{equation}
\makebox[13cm][l]
{
$
\begin{array}{ll@{\hspace{1mm}}c@{\hspace{1mm}}l@{\hspace{1mm}}c@{\hspace{1mm}}l}
\text{\tiny{FR}}&E(y,t|x,s)&=&B(y,t|x,s)-\displaystyle\int _s^td\tau &\displaystyle\frac{\partial}{\partial\tau }&\displaystyle\underset{D}{\int} d\alpha\,E(y,t|\alpha,\tau)B(\alpha,\tau|x,s)\\[2ex]
\text{\tiny{LR}}&E(y,t|x,s)&=&B(y,t|x,s)+\displaystyle\int _s^td\tau &\displaystyle\frac{\partial}{\partial\tau}&\displaystyle\underset{D}{\int } d\alpha\, B(y,t|\alpha,\tau)E(\alpha,\tau|x,s)
\end{array}
$
}
\end{equation}
As usual we may use the PDEs under the integral sign, and use Green's second identity to obtain
\begin{equation}
\makebox[13cm][l]
{
$
\begin{array}{lr@{\hspace{1mm}}c@{\hspace{1mm}}l}
\text{\tiny{FR}}&E(y,t|x,s)&=&B(y,t|x,s)-\displaystyle{\frac{1}{2}\int _s^td\tau \underset{\partial D}{\oint} d\beta\,\,E(y,t|\beta,\tau)\left\{\overleftarrow{\partial_\beta}-\overrightarrow{\partial_\beta}\right\}B(\beta,\tau|x,s)}\\
\text{\tiny{LR}}&E(y,t|x,s)&=&B(y,t|x,s)+\displaystyle{\frac{1}{2}\int _s^td\tau  \underset{\partial D}{\oint} d\beta\,\,B(y,t|\beta,\tau)\left\{\overleftarrow{\partial_\beta}-\overrightarrow{\partial_\beta}\right\}E(\beta,\tau|x,s)}\\
\end{array}
$
}
\end{equation}
Using the boundary conditions, we may write this as
\begin{equation}
\makebox[13cm][l]
{
$
\begin{array}{lr@{\hspace{1mm}}c@{\hspace{1mm}}l}
\text{\tiny{FR}}&E(y,t|x,s)&=&B(y,t|x,s)+\displaystyle{\frac{1}{2}\int _s^td\tau \underset{\partial D}{\oint} d\beta\,\,E(y,t|\beta,\tau)\left\{\overleftarrow{\partial_\beta}+\overrightarrow{\partial_\beta}\right\}B(\beta,\tau|x,s)}\\
&&&\quad\quad\displaystyle- \sigma^2\,\int _s^td\tau \underset{\partial D}{\oint} d\beta\,E(y,t|\beta,\tau)\,\kappa(\beta)\,B(\beta,\tau|x,s)\\
\text{\tiny{LR}}&E(y,t|x,s)&=&B(y,t|x,s)+\displaystyle{\frac{1}{2}\int _s^td\tau  \underset{\partial D}{\oint} d\beta\,\,B(y,t|\beta,\tau)\left\{\overleftarrow{\partial_\beta}+\overrightarrow{\partial_\beta}\right\}E(\beta,\tau|x,s)}\\
&&&\quad\quad\displaystyle- \sigma^2\,\int _s^td\tau \underset{\partial D}{\oint} d\beta\,B(y,t|\beta,\tau)\,\kappa(\beta)\,E(\beta,\tau|x,s)\\
\end{array}
$
}
\end{equation}
Using the divergence theorem, we have
\begin{equation}
\makebox[13cm][l]
{
$
\begin{array}{lr@{\hspace{1mm}}c@{\hspace{1mm}}l}
\text{\tiny{FR}}&E(y,t|x,s)&=&B(y,t|x,s)-\displaystyle \int _s^td\tau \underset{D}{\int} d\alpha\;\frac{\sigma^2}{2}\nabla_\alpha^2\, \bigg[E(y,t|\alpha,\tau)B(\alpha,\tau|x,s)\bigg]\\
&&&\quad\quad\displaystyle- \sigma^2\,\int _s^td\tau \underset{\partial D}{\oint} d\beta\,E(y,t|\beta,\tau)\,\kappa(\beta)\,B(\beta,\tau|x,s)\\
\text{\tiny{LR}}&E(y,t|x,s)&=&B(y,t|x,s)-\displaystyle \int _s^td\tau  \underset{D}{\int} d\alpha\;\frac{\sigma^2}{2}\nabla_\alpha^2\,\bigg[ B(y,t|\alpha,\tau)E(\alpha,\tau|x,s)\bigg]\\
&&&\quad\quad\displaystyle- \sigma^2\,\int _s^td\tau \underset{\partial D}{\oint} d\beta\,B(y,t|\beta,\tau)\,\kappa(\beta)\,E(\beta,\tau|x,s)\\
\end{array}
$
}
\end{equation}
We may use the bump function to write this as 
\begin{equation}
\makebox[13cm][l]
{
$
\begin{array}{lr@{\hspace{1mm}}c@{\hspace{1mm}}l}
\text{\tiny{FR}}&E(y,t|x,s)&=&B(y,t|x,s)-\displaystyle \underset{\epsilon \searrow 0}\lim\int _s^td\tau \underset{\mathbb{R}^d}{\int} d\alpha\; I_\epsilon(\alpha)\frac{\sigma^2}{2}\nabla_\alpha^2\, \bigg[E(y,t|\alpha,\tau)B(\alpha,\tau|x,s)\bigg]\\
&&&\quad\quad\displaystyle- \sigma^2\,\int _s^td\tau \underset{\partial D}{\oint} d\beta\,E(y,t|\beta,\tau)\,\kappa(\beta)\,B(\beta,\tau|x,s)\\
\text{\tiny{LR}}&E(y,t|x,s)&=&B(y,t|x,s)-\displaystyle \underset{\epsilon \searrow 0}\lim\int _s^td\tau  \underset{\mathbb{R}^d}{\int} d\alpha\;I_\epsilon(\alpha)\frac{\sigma^2}{2}\nabla_\alpha^2\,\bigg[ B(y,t|\alpha,\tau)E(\alpha,\tau|x,s)\bigg]\\
&&&\quad\quad\displaystyle- \sigma^2\,\int _s^td\tau \underset{\partial D}{\oint} d\beta\,B(y,t|\beta,\tau)\,\kappa(\beta)\,E(\beta,\tau|x,s)\\
\end{array}
$
}
\end{equation}
where no definition for $E$ is required outside of $D$ (and any definition is allowed), since the indicator is zero there. By Green's theorem where the boundary terms disappear, we obtain 
\begin{equation}
\makebox[13cm][l]
{
$
\begin{array}{lr@{\hspace{1mm}}c@{\hspace{1mm}}l}
\text{\tiny{FR}}&E(y,t|x,s)&=&B(y,t|x,s)-\displaystyle \underset{\epsilon \searrow 0}\lim\int _s^td\tau \underset{\mathbb{R}^d}{\int} d\alpha \;E(y,t|\alpha,\tau)\bigg[\frac{\sigma^2}{2}\nabla_\alpha^2 I_\epsilon(\alpha) \bigg]B(\alpha,\tau|x,s)\\
&&&\quad\quad\displaystyle- \sigma^2\,\int _s^td\tau \underset{\partial D}{\oint} d\beta\,E(y,t|\beta,\tau)\,\kappa(\beta)\,B(\beta,\tau|x,s)\\
\text{\tiny{LR}}&E(y,t|x,s)&=&B(y,t|x,s)-\displaystyle \underset{\epsilon \searrow 0}\lim\int _s^td\tau  \underset{\mathbb{R}^d}{\int} d\alpha\; B(y,t|\alpha,\tau)\bigg[\frac{\sigma^2}{2}\nabla_\alpha^2 I_\epsilon(\alpha)\bigg]E(\alpha,\tau|x,s)\\
&&&\quad\quad\displaystyle- \sigma^2\,\int _s^td\tau \underset{\partial D}{\oint} d\beta\,B(y,t|\beta,\tau)\,\kappa(\beta)\,E(\beta,\tau|x,s)\\
\end{array}
$
}
\end{equation}
Using \eqref{multidimdelta}, we may write this as
\begin{equation}
\makebox[13cm][l]
{
$
\begin{array}{lr@{\hspace{1mm}}c@{\hspace{1mm}}l}
\text{\tiny{FR}}&E(y,t|x,s)&=&B(y,t|x,s)-\\
&&&\hspace{-1cm}\displaystyle \underset{\epsilon \searrow 0}\lim\int _s^td\tau \underset{\mathbb{R}^d}{\int} d\alpha \;E(y,t|\alpha,\tau)\bigg[\frac{\sigma^2}{2}\nabla_\alpha^2 I_\epsilon(\alpha)-\sigma^2\,\kappa(\alpha)\,n_\alpha \cdot\nabla_\alpha I_\epsilon(\alpha) \bigg]B(\alpha,\tau|x,s)\\
\text{\tiny{LR}}&E(y,t|x,s)&=&B(y,t|x,s)-\\
&&&\hspace{-1cm}\displaystyle \underset{\epsilon \searrow 0}\lim\int _s^td\tau  \underset{\mathbb{R}^d}{\int} d\alpha\; B(y,t|\alpha,\tau)\bigg[\frac{\sigma^2}{2}\nabla_\alpha^2 I_\epsilon(\alpha)-\sigma^2\,\kappa(\alpha)\,n_\alpha \cdot\nabla_\alpha I_\epsilon(\alpha)\bigg]E(\alpha,\tau|x,s)\\
\end{array}
$
}
\end{equation}

Comparing with the FI and LI decompositions for singular potentials in Proposition \ref{Proposition6}, we can formally define the elastic potential as follows:
\begin{equation*}
V(\alpha):=\underset{\epsilon \searrow 0}\lim\,\bigg(\frac{\sigma^2}{2}\nabla_\alpha^2 I_\epsilon(\alpha)-\sigma^2\,\kappa(\alpha)\,n_\alpha \cdot\nabla_\alpha I_\epsilon(\alpha)\bigg)
\end{equation*}
where $\kappa(\alpha)$ and $n_\alpha$, which were originally only defined on the boundary, may be defined at any $\alpha$ as being equal to $\kappa(\beta)$ and $n_\beta$, where $\beta$ is the nearest boundary point. 

\subsection{Main theorem}
\label{maintheorem}

We now present the our main theorem:
\begin{Theorem1} 
\label{Theorem1} 
\textnormal{\textbf{Laplacian of the indicator.}}
For all $x,y \in D$ and for all domains $D$ allowing Green's theorem, the absorbed Brownian propagator can be written as the following Feynman-Kac path integral:
\begin{equation*}
\makebox[14cm][l]
{
$A(y,t|x,s)=\underset{\epsilon \searrow 0}\lim\;\mathbb{E}_{x} \Bigg[\,\delta(B_t-y)\, \exp\bigg(\displaystyle \frac{\sigma^2}{2}\int_s^t\,\nabla^2 I_\epsilon(B_{\tau })\,d\tau\bigg)\,\Bigg]\;\mbox{$\forall x,y\in D$}
$
}
\end{equation*}
while the elastic Brownian propagator can be written as:
\begin{equation*}
\makebox[14cm][l]
{
$
E(y,t|x,s)=\underset{\epsilon \searrow 0}\lim\;\mathbb{E}_{x} \Bigg[\,\delta(B_t-y)\, \exp\bigg(\displaystyle- \int_s^t\left(\,\frac{\sigma^2}{2}\nabla^2 I_\epsilon(B_{\tau })-\sigma^2\,\kappa(B_\tau)\,n \cdot\nabla I_\epsilon(B_\tau)\right)\,d\tau\bigg)\,\Bigg]
$
}
\end{equation*}
where the reflected propagator $R$ can be obtained by letting $\kappa \searrow 0$. 
\end{Theorem1}

\begin{proof} 
Recalling the definition of the integral operator $L$ in \eqref{Loperator}, we obtain
\begin{equation}
\begin{array}{r@{\hspace{1mm}}c@{\hspace{1mm}}l}
A(y,t|x,s)&=&\displaystyle B(y,t|x,s)+\sum_{i=1}^\infty\; (-1)^i\, (L*)^i\, B(y,t|x,s)
\end{array}
\end{equation}
where the operator $L$ is defined by \eqref{Loperator}, and where 
\begin{equation}
V_\epsilon(x)=-\frac{\sigma^2}{2} \nabla_x^2 I_\epsilon(x)
\end{equation}
Now consider the first perturbation term, which is $-L*B$. Using the short-hand $I''_\epsilon(x)=\nabla_x^2 I_\epsilon(x)$, we obtain
\begin{equation*}
\begin{array}{r@{\hspace{1mm}}c@{\hspace{1mm}}l}
L*B(y,t|x,s)&=&\displaystyle-\underset{\epsilon \searrow 0}\lim\int _s^td\tau \underset{\mathbb{R}^d}{\int}d\alpha\; B(y,t|\alpha,\tau)\left\{\frac{-\sigma^2}{2} I_\epsilon''(\alpha)\right\}B(\alpha,\tau|x,s)\\
&=&\displaystyle\underset{\epsilon \searrow 0}\lim\int _s^t d\tau \underset{\mathbb{R}^d}{\int}d\alpha\; I_\epsilon(\alpha) \left\{\frac{\sigma^2}{2} \nabla_{{\alpha}}^2 \right\}\bigg[B(y,t|\alpha,\tau)B(\alpha,\tau|x,s)\bigg]\\
&=&\displaystyle\int _s^t d\tau \underset{D}{\int}d\alpha\; \left\{\frac{\sigma^2}{2} \nabla_{{\alpha}}^2 \right\}\bigg[B(y,t|\alpha,\tau)B(\alpha,\tau|x,s)\bigg]\\
&=&- \displaystyle\int _s^td\tau \underset{\partial D}{\int}d\beta\; B(y,t|\beta,\tau)\left\{\frac{1}{2} \overleftarrow{\partial_\beta}+\frac{1}{2} \overrightarrow{\partial_\beta}\right\}B(\beta,\tau|x,s)\\
&=&- \displaystyle\int _s^td\tau \underset{\partial D}{\int}d\beta\; B(y,t|\beta,\tau) \overrightarrow{\partial_\beta}B(\beta,\tau|x,s)
\end{array}
\end{equation*}
where the first equality follows by Green's theorem, the third by the divergence theorem, and the fourth by Lemma \ref{Lemma3}. For a convex domain $D$, the result is negative and indeed equals the first order term in Proposition \ref{Proposition3}. For the second correction term, we need to consider $+L*L*B$, i.e. 
\begin{equation*}
\begin{array}{r@{\hspace{1mm}}c@{\hspace{1mm}}l}
L*L*B(y,t|x,s)&=&\displaystyle\left(\frac{\sigma^2}{2}\right)^2\underset{\epsilon_2 \searrow 0}\lim\int _s^t d\tau_2 \underset{\mathbb{R}^d}{\int}d\alpha_2\; B(y,t|\alpha_2,\tau_2) I_{\epsilon_2}''(\alpha_2)\\
&&\quad\quad\displaystyle\times\underset{\epsilon_1 \searrow 0}\lim\int _s^{\tau_2} d\tau_1 \underset{\mathbb{R}^d}{\int}d\alpha_1\; B(\alpha_2,\tau_2|\alpha_1,\tau_1)I_{\epsilon_1}''(\alpha_1) B(\alpha_1,\tau_1|x,s)
\end{array}
\end{equation*}
where the order of limits and integrations is crucial. By the analysis of the first perturbation term we obtain that
\begin{equation*}
\begin{array}{r@{\hspace{1mm}}c@{\hspace{1mm}}l}
L*L*B(y,t|x,s)&=&\displaystyle-\frac{\sigma^2}{2}\underset{\epsilon \searrow 0}\lim \int _s^t d\tau_2 \underset{\mathbb{R}^d}{\int}d\alpha_2\; B(y,t|\alpha_2,\tau_2)I_{\epsilon}''(\alpha_2)\\
&&\quad\quad\displaystyle\times\int _s^{\tau_2} d\tau_1 \underset{\partial D}{\oint}d\beta_1\; B(\alpha_2,\tau_2|\beta_1,\tau_1)\left\{ \overrightarrow{\partial_{\beta_1}}\right\}B(\beta_1,\tau_1|x,s)
\end{array}
\end{equation*}
By Green's theorem, we obtain
\begin{equation*}
\begin{array}{r@{\hspace{1mm}}c@{\hspace{1mm}}l}
L*L*B(y,t|x,s)&=&\displaystyle-\frac{\sigma^2}{2}\underset{\epsilon \searrow 0}\lim \int _s^t d\tau_2 \underset{\mathbb{R}^d}{\int}d\alpha_2\; I_\epsilon(\alpha_2) \nabla_{\alpha_2}^2 \Bigg[B(y,t|\alpha_2,\tau_2)\\
&&\quad\quad\displaystyle\times\int _s^{\tau_2} d\tau_1 \underset{\partial D}{\oint}d\beta_1\; B(\alpha_2,\tau_2|\beta_1,\tau_1)\left\{ \overrightarrow{\partial_{\beta_1}}\right\}B(\beta_1,\tau_1|x,s)\Bigg]
\end{array}
\end{equation*}
And by taking the limit, we get
\begin{equation*}
\begin{array}{r@{\hspace{1mm}}c@{\hspace{1mm}}l}
L*L*B(y,t|x,s)&=&\displaystyle-\frac{\sigma^2}{2} \int _s^t d\tau_2 \underset{D}{\int}d\alpha_2\; \nabla_{\alpha_2}^2 \Bigg[B(y,t|\alpha_2,\tau_2)\\
&&\quad\quad\displaystyle\times\int _s^{\tau_2} d\tau_1 \underset{\partial D}{\oint}d\beta_1\; B(\alpha_2,\tau_2|\beta_1,\tau_1)\left\{ \overrightarrow{\partial_{\beta_1}}\right\}B(\beta_1,\tau_1|x,s)\Bigg]
\end{array}
\end{equation*}
By the divergence theorem, we then obtain
\begin{equation*}
\begin{array}{r@{\hspace{1mm}}c@{\hspace{1mm}}l}
L*L*B(y,t|x,s)&=&\displaystyle\int _s^t d\tau_2 \underset{\partial D}{\oint}d\beta_2\; B(y,t|\beta_2,\tau_2)\left\{\frac{1}{2} \overleftrightarrow{\partial_{\beta_2}}\right\}\\
&&\quad\quad\displaystyle\times\displaystyle\int _s^{\tau_2} d\tau_1 \underset{\partial D}{\oint}d\beta_1\; B(\beta_2,\tau_2|\beta_1,\tau_1)\left\{ \overrightarrow{\partial_{\beta_1}}\right\}B(\beta_1,\tau_1|x,s)
\end{array}
\end{equation*}
When written out, this becomes
\begin{equation*}
\begin{array}{ll}
&\displaystyle\int _s^t d\tau_2 \underset{\partial D}{\oint}d\beta_2\; B(y,t|\beta_2,\tau_2)\left\{\frac{1}{2} \overleftarrow{\partial_{\beta_2}}\right\}\int _s^{\tau_2} d\tau_1 \underset{\partial D}{\oint}d\beta_1\; B(\beta_2,\tau_2|\beta_1,\tau_1)\left\{ \overrightarrow{\partial_{\beta_1}}\right\}B(\beta_1,\tau_1|x,s)\\
+&\displaystyle\int _s^t d\tau_2 \underset{\partial D}{\oint}d\beta_2\; B(y,t|\beta_2,\tau_2)\left\{\frac{1}{2} \overrightarrow{\partial_{\beta_2}}\right\}\int _s^{\tau_2} d\tau_1 \underset{\partial D}{\oint}d\beta_1\; B(\beta_2,\tau_2|\beta_1,\tau_1)\left\{ \overrightarrow{\partial_{\beta_1}}\right\}B(\beta_1,\tau_1|x,s)
\end{array}
\end{equation*}
By Lemma \ref{Lemma1} we can push the differential operator through the integral in the second term, to obtain
\begin{equation*}
\begin{array}{ll}
&\displaystyle\int _s^t d\tau_2 \underset{\partial D}{\oint}d\beta_2\; B(y,t|\beta_2,\tau_2)\left\{\frac{1}{2} \overleftarrow{\partial_{\beta_2}}\right\}\int _s^{\tau_2} d\tau_1 \underset{\partial D}{\oint}d\beta_1\; B(\beta_2,\tau_2|\beta_1,\tau_1)\left\{ \overrightarrow{\partial_{\beta_1}}\right\}B(\beta_1,\tau_1|x,s)\\
+&\displaystyle\int _s^t d\tau_2 \underset{\partial D}{\oint}d\beta_2\; B(y,t|\beta_2,\tau_2)\int _s^{\tau_2} d\tau_1 \underset{\partial D}{\oint}d\beta_1\; \left\{\frac{1}{2} \overrightarrow{\partial_{\beta_2}}\right\}B(\beta_2,\tau_2|\beta_1,\tau_1)\left\{ \overrightarrow{\partial_{\beta_1}}\right\}B(\beta_1,\tau_1|x,s)\\
-&\displaystyle\int _s^t d\tau_2 \underset{\partial D}{\oint}d\beta_2\; B(y,t|\beta_2,\tau_2)\left\{\frac{1}{2}\overrightarrow{\partial_{\beta_2}}\right\}B(\beta_2,\tau_2|x,s)
\end{array}
\end{equation*}
There are now no more differentiations that are pointing through integral operators, so we may finally pull all the integrals towards the left, to obtain
\begin{equation*}
\begin{array}{ll}
&\displaystyle\int _s^t d\tau_2 \int _s^{\tau_2} d\tau_1\; \underset{\partial D}{\oint}d\beta_2 \underset{\partial D}{\oint}d\beta_1\; B(y,t|\beta_2,\tau_2)\left\{\frac{1}{2} \overleftarrow{\partial_{\beta_2}}\right\}B(\beta_2,\tau_2|\beta_1,\tau_1)\left\{ \overrightarrow{\partial_{\beta_1}}\right\}B(\beta_1,\tau_1|x,s)\\
+&\displaystyle\int _s^t d\tau_2 \int _s^{\tau_2} d\tau_1\; \underset{\partial D}{\oint}d\beta_2 \underset{\partial D}{\oint}d\beta_1\;B(y,t|\beta_2,\tau_2)\left\{\frac{1}{2} \overrightarrow{\partial_{\beta_2}}\right\}B(\beta_2,\tau_2|\beta_1,\tau_1)\left\{ \overrightarrow{\partial_{\beta_1}}\right\}B(\beta_1,\tau_1|x,s)\\
-&\displaystyle\int _s^t d\tau_2 \underset{\partial D}{\oint}d\beta_2\; B(y,t|\beta_2,\tau_2)\left\{\frac{1}{2}\overrightarrow{\partial_{\beta_2}}\right\}B(\beta_2,\tau_2|x,s)
\end{array}
\end{equation*}
Changing the direction of an arrow, in the first term, by taking into account Lemma \ref{Lemma3}, we obtain
\begin{equation*}
\begin{array}{ll}
&\displaystyle\int _s^t d\tau_2 \int _s^{\tau_2} d\tau_1\; \underset{\partial D}{\oint}d\beta_2 \underset{\partial D}{\oint}d\beta_1\; B(y,t|\beta_2,\tau_2)\left\{\frac{1}{2} \overrightarrow{\partial_{\beta_2}}\right\}B(\beta_2,\tau_2|\beta_1,\tau_1)\left\{ \overrightarrow{\partial_{\beta_1}}\right\}B(\beta_1,\tau_1|x,s)\\
+&\displaystyle\int _s^t d\tau_2 \underset{\partial D}{\oint}d\beta_2\; B(y,t|\beta_2,\tau_2)\left\{\frac{1}{2}\overrightarrow{\partial_{\beta_2}}\right\}B(\beta_2,\tau_2|x,s)\\
+&\displaystyle\int _s^t d\tau_2 \int _s^{\tau_2} d\tau_1\; \underset{\partial D}{\oint}d\beta_2 \underset{\partial D}{\oint}d\beta_1\;B(y,t|\beta_2,\tau_2)\left\{\frac{1}{2} \overrightarrow{\partial_{\beta_2}}\right\}B(\beta_2,\tau_2|\beta_1,\tau_1)\left\{ \overrightarrow{\partial_{\beta_1}}\right\}B(\beta_1,\tau_1|x,s)\\
-&\displaystyle\int _s^t d\tau_2 \underset{\partial D}{\oint}d\beta_2\; B(y,t|\beta_2,\tau_2)\left\{\frac{1}{2}\overrightarrow{\partial_{\beta_2}}\right\}B(\beta_2,\tau_2|x,s)
\end{array}
\end{equation*}
Two terms cancel and collecting the other two, we finally get
\begin{equation*}
\begin{array}{l}
L*L*B(y,t|x,s)=\displaystyle\int _s^t d\tau_2 \int _s^{\tau_2} d\tau_1\; \underset{\partial D}{\oint}d\beta_2 \underset{\partial D}{\oint}d\beta_1\; B(y,t|\beta_2,\tau_2) \overrightarrow{\partial_{\beta_2}}B(\beta_2,\tau_2|\beta_1,\tau_1) \overrightarrow{\partial_{\beta_1}}B(\beta_1,\tau_1|x,s)\\
\end{array}
\end{equation*}
This shows that not only the first but also the second perturbation term of Theorem \ref{Theorem1} is equal to the corresponding one in Proposition \ref{Proposition3}. We noticed earlier that the only difference between the absorbed perturbation series in Proposition \ref{Proposition3} and reflected perturbation series in Proposition \ref{Proposition4} is that the \textit{signs} in front of the perturbation terms are different. We now realise that this is simply because the potential generating them differs by a sign. With the same methods as above --- i.e. using Lemma \ref{Lemma1} to push differential operators through integrals, and by using Lemma \ref{Lemma3} to change the direction of arrows --- we can show that all higher order perturbation terms are also equal to those in Propositions \ref{Proposition3} and \ref{Proposition4}. This completes the proof.
\end{proof}

\section{Moving boundaries}
\label{section5}

\subsection{Time-dependent domains with absorbing BCs}

We denote the time-dependent domain by $D(\cdot)$. The domain and its boundary at a specific time $t$ are indicated by $D(t)$ and $\partial{D}(t)$. The absorbed transition density $A(y,t|x,s)$ satisfies the following set of equations:
\begin{equation}
\label{AbsorbedBrownianMotionMovingBoundary}
\begin{array}{rr@{\hspace{2mm}}c@{\hspace{2mm}}lr@{\hspace{1mm}}lr@{\hspace{1mm}}l}
\text{\scriptsize{forward PDE}}&\displaystyle\left(\frac{\partial}{\partial t} - \frac{1}{2}\nabla_y^2\right)A(y,t|x,s)&=&0&x&\in D(s)& y &\in D(t),\\[1.5ex]
\text{\scriptsize{backward PDE}}&\displaystyle\left(\frac{\partial}{\partial s} + \frac{1}{2}\nabla_x^2\right)A(y,t|x,s)&=&0&x&\in D(s)& y &\in D(t),\\[1.5ex]
\text{\scriptsize{forward BC}}&\displaystyle A(\beta,t|x,s)&=&0&x&\in D(s)& \beta &\in \partial D(t),\\[1.5ex]
\text{\scriptsize{backward BC}}&\displaystyle A(y,t|\beta,s)&=&0&\beta &\in \partial D(s)&y&\in D(t) ,\\[1.5ex]
\text{\scriptsize{forward STC}}&\displaystyle\lim_{s \nearrow t} A(y,t|x,s)&=&\delta(|y-x|)&x&\in D(t)& y &\in D(t),\\[1.5ex]
\text{\scriptsize{backward STC}}&\displaystyle\lim_{t \searrow s} A(y,t|x,s)&=&\delta(|y-x|)&x&\in D(s)& y &\in D(s).\\
\end{array}
\end{equation}
where the boundary conditions hold at regular boundary points, and where we assume that the boundary moves with an integrable speed at all times. Writing down the first- and last-passage decompositions and proceeding as in the case of a static domain, we derive a similar pair of integral equations:
\begin{equation}
\makebox[13cm][l]
{
$
\begin{array}{lr@{\hspace{1mm}}c@{\hspace{1mm}}l}
\textnormal{\tiny{FP}}&A(y,t|x,s)&=&B(y,t|x,s)-\displaystyle{\int _s^t d\tau \underset{\partial D(\tau)}{\oint} d\beta\;B(y,t|\beta,\tau)\left\{\frac{1}{2}\overrightarrow{\partial_\beta}\right\}A(\beta,\tau|x,s)},\\
\textnormal{\tiny{LP}}&A(y,t|x,s)&=&B(y,t|x,s)-\displaystyle{\int _s^t d\tau  \underset{\partial D(\tau)}{\oint} d\beta\;A(y,t|\beta,\tau)\left\{\frac{1}{2}\overleftarrow{\partial_\beta}\right\}B(\beta,\tau|x,s)},
\end{array}
$
}
\end{equation}
with the only difference that the integration is now over the time-dependent domain $D(\tau)$. We can rewrite these decompositions as
\begin{equation}
\makebox[13cm][l]
{
$
\begin{array}{lr@{\hspace{1mm}}c@{\hspace{1mm}}l}
\textnormal{\tiny{FP}}&A(y,t|x,s)&=&B(y,t|x,s)+\displaystyle\frac{\sigma^2}{2}\int _s^t d\tau \underset{D(\tau)}{\int} d\alpha\;\nabla_\alpha^2\bigg[ B(y,t|\alpha,\tau)\,A(\alpha,\tau|x,s)\bigg],\\
\textnormal{\tiny{LP}}&A(y,t|x,s)&=&B(y,t|x,s)+\displaystyle\frac{\sigma^2}{2}\int _s^t d\tau \underset{D(\tau)}{\int} d\alpha\;\nabla_\alpha^2\bigg[ A(y,t|\alpha,\tau)\,B(\alpha,\tau|x,s)\bigg].\\
\end{array}
$
}
\end{equation} 
We may substitute these equations back into themselves to obtain a series solution. For a time-dependent potential $V$, the first- and last-interaction decompositions of Proposition \ref{Proposition5} turn into
\begin{equation}
\label{timedependentdecompositions}
\begin{array}{ll}
\textnormal{\tiny{FI}}&\psi_V(y,t|x,s)=\displaystyle B(y,t|x,s)-\int_s^t d\tau \underset{\mathbb{R}^d}{\int } d\alpha\;B(y,t|\alpha,\tau)\,\lambda\,V(\alpha,\tau)\,\psi _V(\alpha ,\tau |x,s),\\[2ex]
\textnormal{\tiny{LI}}&\psi_V(y,t|x,s)=\displaystyle B(y,t|x,s)-\int_s^t d\tau \underset{\mathbb{R}^d}{\int } d\alpha\;\psi_V(y,t|\alpha,\tau)\,\lambda\,V(\alpha,\tau)\,B(\alpha ,\tau |x,s).
\end{array}
\end{equation}
The only difference is that $V$ is now time- as well as space-dependent. Proceeding heuristically as before, we can show that the potential is given by 
\begin{equation*}
\begin{array}{rcl}
V(\alpha,\tau)&=&\displaystyle- \frac{\sigma^2}{2}\,\nabla_{\alpha}^2 \mathbbm{1}_{\alpha\in D(\tau)},\\[2ex] 
&=&\displaystyle-\frac{\sigma^2}{2}\underset{\epsilon \searrow 0}\lim\,\nabla_\alpha^2 I_\epsilon(\alpha,\tau),
\end{array}
\end{equation*}
where $I_\epsilon(\alpha,\tau)$ approaches $\mathbbm{1}_{\alpha\in D(\tau)}$ from below. 

\subsection{Time-dependent domains with reflecting BCs}

Consider now reflected Brownian motion in the time-dependent domain $D(\cdot)$. In e.g. \cite{Burdzy2004}, a set of backward equations can be found. We add a set of forward equations to obtain the full set:
\begin{equation}
\label{ReflectedBrownianMotionMovingDomain}
\begin{array}{rr@{\hspace{2mm}}c@{\hspace{2mm}}lr@{\hspace{1mm}}lr@{\hspace{1mm}}l}
\text{\scriptsize{forward PDE}}&\displaystyle\left(\frac{\partial}{\partial t} - \frac{1}{2}\nabla_y^2\right)R(y,t|x,s)&=&0&x&\in D(s)& y &\in D(t),\\[1.5ex]
\text{\scriptsize{backward PDE}}&\displaystyle\left(\frac{\partial}{\partial s} + \frac{1}{2}\nabla_x^2\right)R(y,t|x,s)&=&0&x&\in D(s)& y &\in D(t),\\[1.5ex]
\text{\scriptsize{forward BC}} &\displaystyle\left(\overrightarrow{\partial_\beta}-2\dot{\beta}(t)\cdot n_\beta(t)\right) R(\beta,t|x,s)&=&0&x&\in D(s)& \beta &\in \partial D(t),\\[1.5ex]
\text{\scriptsize{backward BC}}&\displaystyle R(y,t|\beta,s)\overleftarrow{\partial_\beta}&=&0&\beta &\in \partial D(s)&y&\in D(t), \\[1.5ex]
\text{\scriptsize{forward STC}}&\displaystyle\lim_{s \nearrow t} R(y,t|x,s)&=&\delta(|y-x|)&x&\in D(t)& y &\in D(t),\\[1.5ex]
\text{\scriptsize{backward STC}}&\displaystyle\lim_{t \searrow s} R(y,t|x,s)&=&\delta(|y-x|)&x&\in D(s)& y &\in D(s).
\end{array}
\end{equation}
The backward BC is unchanged from the time-independent case, but the forward BC may be surprising. It can be derived as follows. By Chapman-Kolmogorov, we have
\begin{equation}
R(y,t|x,s)=\displaystyle\underset{D(\tau)}{\int} d\alpha\,R(y,t|\alpha,\tau)R(\alpha,\tau|x,s),
\end{equation}
for $s\leq\tau\leq t$. The left-hand side does not depend on $\tau$. Differentiating with respect to $\tau$, using Reynold's transport theorem (as in \cite{Lorenz2006}), and integrating under the integral sign gives
\begin{equation}
\begin{array}{rcl}
0&=&\displaystyle\frac{\partial}{\partial \tau}\underset{D(\tau)}{\int} d\alpha\,R(y,t|\alpha,\tau)R(\alpha,\tau|x,s),\\
0&=&\displaystyle\underset{\partial D(\tau)}{\oint} d\beta\, R(y,t|\beta,\tau)\left\{n_\beta\cdot \dot{\beta}(\tau)\right\}R(\beta,\tau|x,s)\\
&&\quad-\displaystyle{\frac{1}{2} \underset{D(\tau)}{\int} d\alpha\,\,R(y,t|\alpha,\tau)\left \{\overleftarrow\nabla_\alpha^2-\overrightarrow\nabla_\alpha^2\right\}R(\alpha,\tau|x,s)},\\
0&=&\displaystyle\underset{\partial D(\tau)}{\oint} d\beta\, R(y,t|\beta,\tau)\left\{n_\beta\cdot \dot{\beta}(\tau)\right\}R(\beta,\tau|x,s)\\
&&\quad\displaystyle{+\frac{1}{2} \underset{\partial D(\tau)}{\oint} d\beta\,\,R(y,t|\beta,\tau)\left\{\overleftarrow{\partial_\beta}-\overrightarrow{\partial_\beta}\right\}R(\beta,\tau|x,s)},
\end{array}
\end{equation}
where $\dot\beta(\tau)$ is the velocity-vector of the boundary element $\beta(\tau)$ and $n_\beta$ is the outward normal. Reynold's transport theorem requires that the boundary moves with integrable speed, which we assume. The last equality follows from Green's theorem, which holds as long as $D(\tau)$ is piecewise smooth for each $\tau$. Using the backward BC, and given that this should hold for \textit{each} domain $D(\cdot)$, it follows that the forward BC must hold at each boundary location $\beta$. 
Now that we have established the BCs, we can write down the first- and last reflection decompositions as in the static case, turn the crank once more, and obtain finally: 
\begin{equation}
\label{ReflectedIntegralEquationsMovingDomain}
\makebox[13cm][l]
{
$
\begin{array}{lr@{\hspace{1mm}}c@{\hspace{1mm}}l}
\textnormal{\tiny{FR}}&R(y,t|x,s)&=&B(y,t|x,s)+\displaystyle\int _s^td\tau \underset{\partial D(\tau)}{\oint} d\beta\;R(y,t|\beta,\tau)\left\{-n_\beta\cdot \dot{\beta}(\tau)+\frac{1}{2}\overrightarrow{\partial_\beta}\right\}B(\beta,\tau|x,s),\\
\textnormal{\tiny{LR}}&R(y,t|x,s)&=&B(y,t|x,s)+\displaystyle\int _s^td\tau  \underset{\partial D(\tau)}{\oint} d\beta\;B(y,t|\beta,\tau)\left\{\frac{1}{2}\overleftarrow{\partial_\beta}\right\}R(\beta,\tau|x,s).\\
\end{array}
$
}
\end{equation}
We may use these decompositions to obtain two series solutions, as in the static case. Or, we may rewrite them to obtain the following more `symmetric' pair:
\begin{equation}
\makebox[13cm][l]
{
$
\begin{array}{lr@{\hspace{1mm}}c@{\hspace{1mm}}l}
\textnormal{\tiny{FR}}&R(y,t|x,s)&=&B(y,t|x,s)-\displaystyle\frac{\sigma^2}{2}\int _s^td\tau \underset{D(\tau)}{\int} d\alpha\;\nabla_\alpha^2 \bigg[R(y,t|\alpha,\tau)\,B(\alpha,\tau|x,s)\bigg]\\
&&&\quad-\displaystyle\int _s^td\tau  \underset{\partial D(\tau)}{\oint} d\beta\;R(y,t|\beta,\tau)\,n_\beta\cdot \dot{\beta}(\tau)\,B(\beta,\tau|x,s)\\
\textnormal{\tiny{LR}}&R(y,t|x,s)&=&B(y,t|x,s)-\displaystyle\frac{\sigma^2}{2}\int _s^td\tau \underset{D(\tau)}{\int} d\alpha\;\nabla_\alpha^2 \bigg[B(y,t|\alpha,\tau)\,R(\alpha,\tau|x,s)\bigg]\\
&&&\quad-\displaystyle\int _s^td\tau  \underset{\partial D(\tau)}{\oint} d\beta\;B(y,t|\beta,\tau)\,n_\beta\cdot \dot{\beta}(\tau)\,R(\beta,\tau|x,s).
\end{array}
$
}
\end{equation}
and thus we may associate the potential 
\begin{equation*}
\begin{array}{rcl}
V(x,\tau)&=&\displaystyle \frac{\sigma^2}{2}\,\nabla_{x}^2 \mathbbm{1}_{x\in D(\tau)}+\dot{\beta}(x)\cdot n_x\,n_x \cdot\nabla_x I_\epsilon(x)
\end{array}
\end{equation*}
where $\dot{\beta}(x)$ and $n_x$ are the vector fields defined by the velocity and normal vectors  of the nearest boundary point, respectively.
\section{Conclusion}
\label{section6}

This paper has considered the heat equation with boundary conditions (section \ref{section2}), and the Schr\"{o}dinger equation with (a possibly singular) potential $V$ (section \ref{section3}). The first- and last-passage decompositions of section \ref{section2} are very similar to the first- and last-interaction decompositions of section \ref{section3}. Section \ref{section4} showed that sections \ref{section2} and \ref{section3} are not merely similar, but equivalent, if we postulate a particular singular potential $V$ (Theorem \ref{Theorem1}).

Section \ref{section2} developed the first- and last-passage (and reflection) decompositions,  showing the equivalence of the \textit{path decomposition expansion} (for path integrals), the \textit{multiple reflection expansion} (found in potential theory) and the series solution of the \textit{parametrix method} (found in probability theory). In particular, 1)  \textit{single} and \textit{double boundary layers} need not be based on an \textit{ansatz}, but follow from the first- and last-interaction decompositions of a Brownian motion, 2) either the absorbed or the reflected propagator may be found with either method, and 3) boundary layers may be useful for irregular as well as regular domains, by virtue of Green's theorem.

Section \ref{section3}, in an analogous manner to section \ref{section2}, developed the first- and last-interaction decompositions for smooth and singular potentials in $\mathbb{R}^d$. It showed how a series solution for a singular potential can be obtained, allowing e.g. \textit{point interactions} and \textit{surface interactions}. 

Section \ref{section4}, finally, postulated a potential that has the (scaled) Laplacian of the indicator as its limit (when $\kappa$ is zero). The Laplacian of the indicator has --- to the author's best knowledge --- not formally been defined before. It can be defined in the theory of distributions by two partial integrations under the integral sign, or by a limiting procedure involving a \textit{bump function}. The potential shows, for the first time, that the Dirichlet and Neumann problems are very closely related: the potential generating the absorbed/reflected density differs only by a \textit{sign}. This also explains why the perturbation series for both problems have different signs for odd-numbered terms, as has been noted in the literature. 

In terms of the intuition for this potential, we have noted that positive potentials destroy particles while negative potentials create particles. From a limiting procedure, as illustrated in Figure \ref{castle}, we can derive intuitively from which side the boundary is reflecting and absorbing. 

We differ from the literature on point interactions in that we consider the higher-dimensional analogue of the the Dirac $\delta$- and $\delta'$-functions to be $-n\cdot\nabla_x\mathbbm{1}_{{x}\in D}$ and $\nabla_x^2 \mathbbm{1}_{x \in D}$, respectively. Both quantities are supported by a \textit{surface} instead of by a \textit{point}. This generalisation is useful because surface-interactions can lead to boundary conditions in $d \geq 1$, while point interactions cannot. 

For moving boundaries many of the same methods are appropriate, as we have shown in section \ref{section5}.

In conclusion, this paper has introduced the Laplacian of the indicator as the crucial tool to write the heat kernel with boundary conditions as a path integral. If one were aiming to communicate this heat kernel in the least possible number of bits, then the solution provided in this paper would be a good candidate. 

\section*{Acknowledgements}

This paper is a shortened (in length) and extended (in content) version of Part I of my
PhD thesis at King's College, University of Cambridge. I would very much like to thank
my supervisor, Professor Daniel Ralph, for his support over the years. I would also like
to thank professors Richard Weber and Kevin Glazebrook. Further thanks are due to the
Electricity Policy Research Group (EPRG) for its financial support, and in particular to
Dr Karsten Neuhoff, Dr Pierre Noel and Professor David Newbery of the EPRG. Lastly, I
would like to thank the anonymous reviewer of this paper for suggestions, which I took up
directly.


\bibliographystyle{./bibliography/JHEP}
\bibliography{./bibliography/bib}

\end{document}